\documentclass[10pt,journal,compsoc]{IEEEtran}

\IEEEoverridecommandlockouts

\ifCLASSOPTIONcompsoc
  \usepackage[nocompress]{cite}
\else
  \usepackage{cite}
\fi

\usepackage{amsmath,amssymb,amsfonts}
\ifCLASSINFOpdf
  \usepackage[pdftex]{graphicx}
  \DeclareGraphicsExtensions{.pdf,.jpeg,.jpg,.png}
\else
  \usepackage[dvips]{graphicx}
  \DeclareGraphicsExtensions{.eps}
\fi

\usepackage{tabu}
\usepackage{booktabs}
\usepackage{multirow}

\usepackage{graphicx}
\usepackage{textcomp}
\usepackage{xcolor}

\usepackage{latexcolors}

\usepackage[noline,linesnumbered,ruled]{algorithm2e}
\SetKw{Payload}{payload}
\SetKw{Initial}{initial}
\SetKw{Let}{let}
\SetKw{Pre}{pre}
\SetKwBlock{Query}{query}{}
\SetKwBlock{Update}{update}{}
\SetKwBlock{AtSource}{prepare}{}
\SetKwBlock{DownStream}{effect}{}
\SetKwComment{tcp}{$\vartriangleright$ }{}

\SetCommentSty{mycommfont}
\SetAlgorithmName{Algorithm}{Algorithm}

\newcommand{\mc}[1]{\mathcal{#1}}
\newcommand{\set}[1]{\{#1\}}
\newcommand{\po}{\stackrel{po}{\longrightarrow}}
\newcommand{\lr}{\stackrel{lr}{\longrightarrow}}
\newcommand{\vis}{\stackrel{vis}{\longrightarrow}}

\newcommand{\pfend}{\hfill $\Box$}

\newcommand{\nordis}[2]{$\mathcal{N}(#1, #2)$}

\def\frcolor{blue}
\def\frm#1{\textcolor{\frcolor}{$\langle$#1$\rangle$}}

\usepackage{ntheorem}
\theoremheaderfont{\bfseries\itshape}
\theorembodyfont{\upshape\parindent0pt}
\theoremseparator{.}
\newtheorem{mydef}{Definition}
\newenvironment{definition}{\begin{mydef}}{\pfend\end{mydef}}

\newcommand{\rwf}{\textsc{Rwf}}
\newcommand{\rwff}{\textsc{Rwf }}

\begin{document}

\title{Remove-Win: a Design Framework for \\ Conflict-free Replicated Data Types}

\author{Yuqi~Zhang,
        Hengfeng~Wei,
        and~Yu~Huang
\IEEEcompsocitemizethanks{\IEEEcompsocthanksitem Yuqi Zhang, Hengfeng Wei, and Yu Huang are with the State Key Laboratory for Novel Software Technology, and Department of Computer Science and Technology, Nanjing University, China, 210023.	\protect\\
E-mail: cs.yqzhang@gmail.com, \{hfwei, yuhuang\}@nju.edu.cn}
\thanks{(Corresponding author: Hengfeng Wei and Yu Huang.)}}

\IEEEtitleabstractindextext{

\begin{abstract}

Distributed storage systems employ replication to improve performance and reliability. To provide low latency data access, replicas are often required to accept updates without coordination with each other, and the updates are then propagated asynchronously. This brings the critical challenge of conflict resolution among concurrent updates. Conflict-free Replicated Data Type (CRDT) is a principled approach to addressing this challenge. However, existing CRDT designs are tricky, and hard to be generalized to other data types. A design framework is in great need to guide the systematic design of new CRDTs.

To address this challenge, we propose \rwff -- the \underline{R}emove-\underline{W}in design \underline{F}ramework for CRDTs. \rwff leverages the simple but powerful remove-win strategy to resolve conflicting updates, and provides generic design for a variety of data container types. Two exemplar implementations following \rwff are given over the Redis data type store, which demonstrate the effectiveness of \rwf. Performance measurements of our implementations further show the efficiency of CRDT designs following \rwf.

\end{abstract}

\begin{IEEEkeywords}
    CRDT, remove-win, replicated data store
\end{IEEEkeywords}
}

\maketitle



\section{Introduction}

Internet-scale distributed systems often replicate application state and logic to reduce user-perceived latency and improve application throughput, while tolerating partial failures \cite{Shapiro11b, Preguica18}. 
In such distributed systems, user-perceived latency and overall service availability are widely regarded as the most critical factors for a large class of applications. 
Thus, many Internet-scale distributed systems are designed for low latency and high availability in the first place \cite{Lloyd13, Gondelman21}. 
To provide low latency and high availability, the update requests must be handled immediately, without communication with remote replicas. Updates can only be asynchronously transmitted to remote replicas, and rolling-back updates to handle conflicts is not acceptable. 

According to the CAP theorem, low latency and high availability can only be achieved at the cost of accepting weak consistency \cite{Brewer00, Gilbert12}. To provide certain guarantees to developers of upper-layer applications, \textit{eventual convergence} is widely accepted, which ensures that when any two replicas have received the same set of updates, they reach the same state \cite{Shapiro11a}. Eventually consistent replicated data types are widely used in scenarios where responsiveness is critical, e.g. in collaborative editing \cite{Wei18}, distributed caching \cite{Bussey19} or coordination-avoidance in databases \cite{Bailis14}. The design of replicated data types guaranteeing eventual convergence brings the challenge of conflict resolution for concurrent updates on different replicas of logically the same data element. The Conflict-free Replicated Data Type (CRDT) framework provides a principled approach to addressing this challenge \cite{Shapiro11b, Preguica18}.

The conflict resolution is typically hard and error-prone, especially for data types having complex semantics. This explains why existing CRDT designs are tricky, and why it is hard to generalize design for one type to other similar types \cite{Shapiro11b, Preguica18}. A design framework is in great need to guide the systematic design of new CRDTs, and the design of CRDTs needs to shift from a craft to an engineering discipline. The essential issue of proposing a design framework is to refine the commonalities among different CRDT designs. Thus the developer can focus on designing special features pertinent to each data type and reuse the common design based on the framework. In this way, the design framework can help even not-experienced developers handle complex and error-prone CRDT designs.

Toward this objective, we propose \rwff -- the \underline{R}emove-\underline{W}in design \underline{F}ramework for CRDTs. \rwff aims at facilitating the design of replicated data container types. A data container is first a set of unique data elements. Existence of each element is identified by its \textit{key}. Moreover, each data element can have \textit{values}. Complex semantics of the data type and the structure among the data elements are ``encoded" in the values of the data elements.

\rwff facilitates the design of replicated data container types leveraging the simple but powerful remove-win strategy for conflict resolution. The basic rationale of the remove-win strategy is that when any operation is concurrent with a remove operation, the remove operation wins. This means that the data element involved in the operations will be eliminated from the container. One salient feature of the remove-win strategy is that, it is independent of  the semantics of the data type under concern. The remove operation simply eliminate the data element, no matter how complex the semantics of the data type are. Though elimination of one element may affect the overall structure of the data container, the maintenance of the structure of the container is independently handled by each replica and requires no coordination with remote peer replicas. 
The salient feature of the remove-win strategy makes it applicable to different data container types and one design framework is proposed to capture the common remove-win resolution for different data types.

Note that the remove-win strategy adopted in \rwff is different from the remove-win strategy used in the existing work, e.g. in the Remove-Win Set\cite{Zaw15}. When a non-remove operation is concurrent with a remove operation, the remove-win strategy in the existing work makes all replicas put the remove operation behind the non-remove operation.  Thus the effect of the remove operation will overwrite that of the preceding operations. In the remove-win strategy used in \rwf, the data element is simply eliminated, requiring no further processing. Our strategy is more simple but also more powerful. It can be more easily applied to different data types.

\rwff provides a generic algorithm skeleton for conflict-free replicated data container types (denoted as \rwf-DTs). User-defined logics are implemented as stubs and inserted
into the skeleton to obtain concrete \rwf-DT designs. The \rwff framework can be implemented over different data type stores. We present an exemplar implementation over the widely used Redis data type store. In the implementation level, \rwff provides a template for \rwf-DT implementations. Common logics of CRDTs as well as those of \rwf-DTs are provided in the template. The user only needs to provide logics pertinent to the specific data type under development.

The usefulness of \rwff is illustrated by two exemplar \rwf-DT implementations -- implementation of a priority queue and that of a list. Performance measurements of our implementations also show the efficiency of CRDT designs following \rwf.

The rest of this work is organized as follows. In Section \ref{Sec: Overview}, we overview our design framework. In Section \ref{Sec: Design} and \ref{Sec: Impl}, we present the generic design of \rwf-DTs and provide an exemplar implementation. Section \ref{Sec: Exp} presents the performance evaluation results. Section \ref{Sec: RW} discusses the related work. In Section \ref{Sec: Concl}, we summarize our work and discuss the future work.

\section{\rwff Overview} \label{Sec: Overview}

The \rwff design framework first decomposes the design of \rwf-DTs into two dimensions. It then provides a template for \rwf-DT implementations, as detailed below.


\subsection{Design of \rwf-DTs}

\begin{figure}[htbp]
    \centering
    \includegraphics[width=0.95\linewidth]{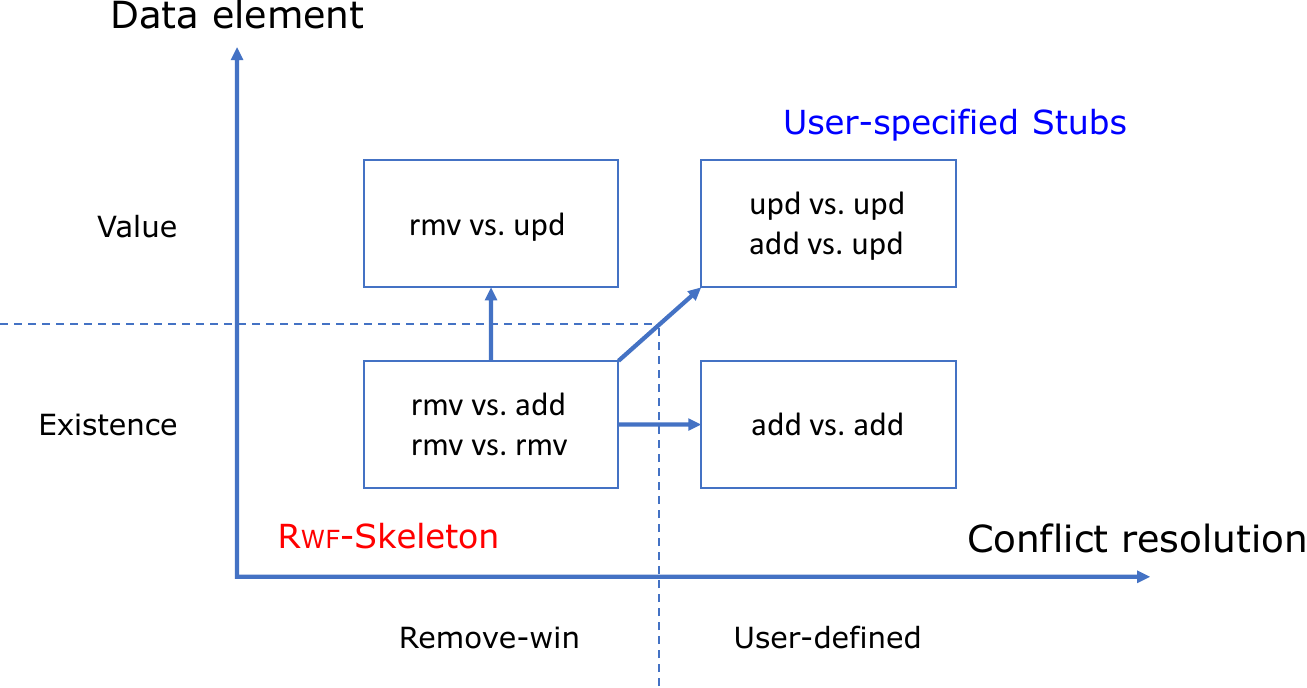}
    \caption{Two dimensions in \rwf-DT design.}
    \label{F: 2D-Framework}
\end{figure}

The \rwff design framework refines the commonalities in CRDT design from two dimensions, as shown in Fig. \ref{F: 2D-Framework}. \rwff first extracts the commonalities from different data types. \rwff focuses on the data container types. Each element in the container first has its unique existence, which is modified by the $add$ and $rmv$ operations. Each data element can also be associated with values, which is modified by the $upd$ operation\footnote{\scriptsize Possibly a container type can have multiple $upd$ operations. We mention only one $upd$ operation for the ease of presentation. Also we only consider ``pure" operations, i.e. each operation is either a query or an update.}. 
Elements in the container may collectively form complex data structures, such as lists, queues and trees. The data structure info is encoded in the value of each element.


\rwff employs the \textit{remove-win} strategy to resolve conflicts between concurrent updates. For conflicting updates involving a $rmv$ and a non-remove operation (i.e., $add$ or $upd$), the $rmv$ operation just eliminates the existence of the data element, no matter what value the element has. For non-remove operations, \rwff requires the user provide conflict resolution logics. The remove-win strategy common to different \rwf-DTs is implemented in an \textit{algorithm skeleton}. User-specified conflict resolution logics are implemented as \textit{stubs}, which can be inserted into the skeleton to obtain concrete \rwf-DT designs, as detailed in the following Section \ref{Sec: Design}.




\subsection{Implementation of \rwf-DTs}

\begin{figure}[htbp]
    \centering
    \includegraphics[width=0.7\linewidth]{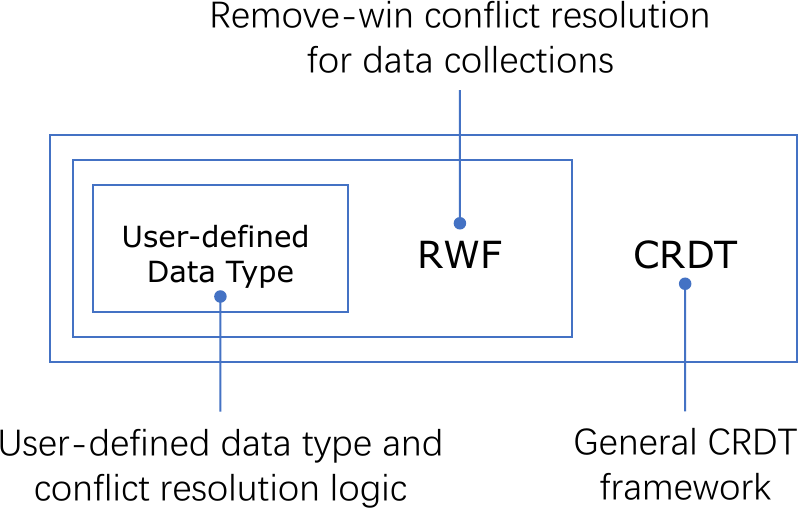}
    \caption{Three layers in the \rwf-DT implementation.}
    \label{F: 3Level-Framework}
\end{figure}

Based on the commonalities in the design, \rwff further provides a template for \rwf-DT implementations, as shown in Fig. \ref{F: 3Level-Framework}. The template has the ``onion" structure and consists of three levels, namely the CRDT level, the \rwff level and the user-defined data type level (denoted as the DT level in short).

In the CRDT level, the basic structure of the implementation is decided, following the operation-based CRDT algorithm framework\cite{Shapiro11a}. Common operations required by the CRDT framework are implemented as tool functions/macros and can be reused for different \rwf-DTs.

In the \rwff level, common metadata pertinent to the predetermined remove-win strategy is defined. Common operations pertinent to the remove-win strategy are also implemented as tool functions. 

In both the CRDT level and the \rwff level, tool functions contain logics which are generic and independent of the specific type of data element in the data container. The user only needs to pass specific type of the data element to the tool functions in the DT level. Moreover, the user also needs to provide conflict-resolution logics which can only be decided by the users.

\section{\rwf-DT Design} \label{Sec: Design}

In this section, we first describe the system model. Then we present design of the \rwf-Set, which is the core of \rwf-DT design. Finally, an algorithm skeleton is presented.

\subsection{System Model}

We use the typical system model for CRDT \cite{Shapiro11b}. Suppose there are $n$ server processes $p_0, p_1, \cdots , p_{n-1}$, each holding one replica of an \rwf-DT. 
Servers are interconnected by an asynchronous network, and can only fail by crash. Messages may be delayed, reordered but cannot be forged. The communication network ensures that eventually all messages are delivered successfully.

\subsubsection{Temporal Order among Events and Operations}

One update operation $o$ initiated on $p_i$ consists of one local event $o.e_{l}$ on $p_i$, and $n$ remote events, one remote event $o.e_{r}$ for each replica, including $p_i$ itself\footnote{\scriptsize For the ease of presentation, the remote event on the initiating process is omitted.}. Here, we say the operation $o$ has executed on replica $p_i$ at time $t$, denoted by $o\in E(p_i^t)$ where $p_i^t$ is the replica state of $p_i$ at time $t$\footnote{\scriptsize We use $p^{cur}$ to denote the current state of replica $p$.}, and $E(p_i^t)$ is the set of executed operations of $p_i^t$, if $o.e_l$ or any of $o.e_r$ has taken place on $p_i$. We define function $\text{TYPE}(o)$, which maps operation $o$ to its type (e.g, $add$, $rmv$ or $upd$).

The temporal order among local and remote events are essential to the design of \rwf-DTs:

\begin{definition}[order between events]
    There are two basic types of order between events:
\begin{itemize}
    \item \textit{Program order}. Events on the same replica are totally ordered by the program order, denoted by $\po$.
    \item \textit{Local-remote order}. The local event $o.e_l$ and each remote event $o.e_r$ belonging to the same operation $o$ have the local-remote order, denoted by $\lr$.
\end{itemize}
    The \textit{happen-before relation} between events, denoted by $\rightarrow$, is defined as the transitive closure of the program order and the local-remote order.
\end{definition}

Given the order between events, we can further define the \textit{visibility}  relation between operations:




\begin{definition}[visibility]
    Operation $o_1$ is visible to $o_2$, denoted by $o_1\vis o_2$, if $o_1.e_l \rightarrow o_2.e_l$. Operation $o$ is visible to replica state $p^t$, if $o\in E(p^t)\vee \exists o':o'\in E(p^t)\wedge o\vis o'$.
\end{definition}

Note that the $\vis$ relation is transitive. Two update operations $o_1$ and $o_2$ are concurrent, denoted by $o_1\parallel o_2$, if neither $o_1\vis o_2$ nor $o_2\vis o_1$ holds.

The importance of the $\vis$ relation is obvious. The remove-win strategy is interpreted with the $\vis$ relation as: non-remove operations which are visible to or are concurrent with a remove operation is eliminated by this remove operation. 

\subsubsection{Segmenting System Execution into Phases}

Given the remove-win strategy, the execution is segmented into phases. Within a phase, non-remove operations initialize a data item and update its value. The remove operation wipes off everything and ends the current phase, and then starts a new phase from scratch. Phase-based resolution is central to the design of \rwf-DTs, as detailed below.

%
%

To define the concept of phase, we first define the remove history of an operation and a replica state:

\begin{definition}[remove history]
    The remove history $\mc{H}_r(o)$ of an operation $o$ is the set of all remove operations that are visible to it: 
\begin{equation*}
    \mc{H}_r(o)=\set{op \ | \ \text{TYPE}(op) = rmv,\ op \vis o}
\end{equation*} 
    The remove history $\mc{H}_r(p^t)$ of one replica state $p^t$ is defined as the union of remove histories of all operations executed on this replica, together with all the remove operations executed on this replica:
    \begin{equation*}
        \mc{H}_r(p^t)=\cup_{o \in E(p^t)}\mc{H}_r(o)\cup \set{o \ | \ \text{TYPE}(o) = rmv,\ o \in E(p^t)}
    \end{equation*} 
\end{definition}

Note that $\mc{H}_r(o)$ is defined for both non-remove and remove operations.

With the definition of remove history, we can formally define \textit{phase}:

\begin{definition}[phase]
    Operations and replica states belong to the same phase, if they have the same remove history. Or equivalently, the phases of the system execution are the equivalence classes in $(O\cup S)/\approx_{\mc{H}_r}$, where $O$ is the set of operations, $S$ is the set of replica states, and $\approx_{\mc{H}_r}$ is the equivalence relation defined by $\mc{H}_r(\cdot)$: 
\begin{equation*}
    a\approx_{\mc{H}_r}b \triangleq \mc{H}_r(a)=\mc{H}_r(b).
\end{equation*}
We denote the phase that operation/replica state $a$ belongs to as $[a]$.
\end{definition}

Phases are temporally ordered. We say $[a]\prec [b]$, if $\mc{H}_r(a)\subset\mc{H}_r(b)$. 

%
%
%

\subsection{Design of the \rwf-Set}

Given the definition of $\vis$ and $\mc{H}_r(\cdot)$, we can now present the design of an \rwf-DT. For the ease of presentation, we first present the core of the design, which is the design of an \rwf-Set. Then we augment the design of the \rwf-Set into an algorithm skeleton, which greatly simplifies the design of various replicated data container types.

\subsubsection{Encoding of Remove History}

Since our design is centered around the remove history, we first discuss how to efficiently encode the remove history for each operation. The remove operation has the salient feature that it does not require any parameters (except for $e$ identifying the element of concern), it is idempotent and its effect is always the same (wiping off everything) no matter how the value of the data element has changed. Thus we do not care how many times the remove operations have taken place. If the $k^{th}$ remove operation that is initiated by $p_i$ is visible, all remove operations, from the $1^{st}$ to the $(k-1)^{th}$, initiated by $p_i$ are visible as well. Since the remove operation is idempotent, we only need to record the last remove operation initiated on $p_i$. 

The encoding/decoding scheme we use is principally the vector clock \cite{Mattern89}. The remove operations visible to an operation $o$ or some replica state $p^t$ can be encoded as a vector $v[1..n]$, which we call the remove history vector (abbreviated as rh-vec). All remove operations initiated on replica $p_i$ are totally ordered, and we use the index $k$ to uniquely identify each remove operation. When we have $v[j]=k$ on replica $p_i$, it means that the last remove operation initiated by $p_j$ that is visible to $p_i^{cur}$ is $p_j$'s $k^{th}$ remove operation (remove operations visible to an operation $o$ is defined similarly).
When replica $p_i$ receives an operation $o$ carrying a rh-vec $v[1..n]$, 
$p_i$'s local rh-vec $v_i[1..n]$ 
needs to be updated as: $\forall j \in [1..n]:v_i[j]=\max(t_i[j],t[j])$. 

\subsubsection{Payload of an \rwf-Set}

Following the CRDT framework, each \rwf-Set $\mc{S}$ is implemented over its payload, two sets $E$ and $T$. On one replica of $\mc{S}$, set $E$ contains the IDs of data elements. Element $e \in E$ basically means that this element is in $\mc{S}$. Set $T$ is the set of tuples $(e, t)$, where tag $t$ is the rh-vec encoding the remove history of the current replica state, concerning data element $e$.

We first discuss how $add$ and $rmv$ operations update the payload. When an $add$ operation $add(e)$ is initiated on replica $p_i$, it first conducts the local processing, taking $e$ as the user-specified parameter (the \textsf{prepare} part, Line 4 -- 7 in Algorithm \ref{A: RWF-Skeleton}\footnote{\scriptsize The Algorithm \ref{A: RWF-Skeleton} contains the \rwf-Set Algorithm, with some detailed extensions like more parameters/steps.}). Replica $p_i$ checks whether $e$ is already in $\mc{S}$ (Line 5). If not, the remove history of this add operation is obtained as $v^{rh}$ (Line 6). After the local processing on the initiating replica $p_i$, $p_i$ broadcasts this $add(e)$ operation and triggers the remote processing on all replicas (the \textsf{effect} part, Line 8 -- 13 in Algorithm \ref{A: RWF-Skeleton}). This broadcast has two parameters, the user-specified parameter $e$ and the parameter $v^{rh}$ prepared in the local processing.

For a remove operation $rmv(e)$, the initiating replica $p_{ini}$ first checks whether this element is actually in $\mc{S}$, and then it locally increases the rh-vec $t[p_{ini}]$ to record this remove operation (Line 19 in Algorithm \ref{A: RWF-Skeleton}). The remove history of this operation is prepared in $v^{rh}$ for the broadcast (Line 17). The user-specified parameter $e$ and locally prepared parameter $v^{rh}$ are broadcast to remote replicas on behalf of the operation $rmv(e)$. If in any dimension $k$, the local rh-vec element $t[k]$ is older than the vector element $v^{rh}[k]$ from the broadcast, we remove $e$ from $E$, since there are unseen remove operations (Line 22--24). Then the local rh-vec $t[1..n]$ is updated to the pairwise maximum of $v^{rh}$ and $t$, and this update is recorded in the payload $T$ (Line 25 -- 26).

\subsubsection{Conflict Resolution for \rwf-Set} \label{SubSubSec: Conf-Res}

To resolve the conflict between concurrent operations, we first need to handle the anomaly caused by the fact that the remove operation can arrive at the remote replica arbitrarily late, since we do not require the communication channel provide causal message delivery \cite{Birman91}. 
This means that when an $add(e)$ operation arrives at $p_i$, the $rmv(e)$ operations visible to it may have not arrived yet. This means that the phase of $p_i^{cur}$ may precede the phase of $add(e)$. However, since all the $rmv(e)$ do not need additional parameters, and the rh-vec $v^{rh}$ of $add(e)$ encodes all the visible $rmv(e)$, we can do these missing $rmv(e)$ operations first (Line 9 in Algorithm \ref{A: RWF-Skeleton}), update the remove history of $p_i^{cur}$, and then do the $add(e)$ operation.

We now discuss the conflict resolution between concurrent $add$ and $rmv$ operations. Suppose operation $add(e)$ is initiated at replica $p_i$. Then the remote event of $add(e)$ arrives at a remote replica $p_j$. Note that the remote event from $p_i$ brings with it the remove history $v^{rh}$ of the $add(e)$ operation (Line 8 in Algorithm \ref{A: RWF-Skeleton}). The rh-vec on remote replica $p_j$ is recorded in its local payload $T$, denoted as $t$. With the supplement of missing $rmv(e)$ operations, $t$ has been updated by $v^{rh}$. We now have $v^{rh} \leq t$. Given this fact, we have two cases left to handle:
\begin{itemize}
    \item $v^{rh} = t$. This means that $add(e)$ and $p_j^{cur}$ have seen the same set of remove operations. There will be no conflict, and we directly add $e$ into payload $E$ on $p_j$.
    
    \item  $v^{rh} < t$. This means that $\exists rmv(e): rmv(e) \vis p_j^{cur} \wedge \lnot(rmv(e) \vis add(e))$. This $rmv(e)$ either is concurrent with $add(e)$ or happens after $add(e)$. According to the remove-win strategy, the effect of $add(e)$ will be wiped off by $rmv(e)$.
\end{itemize}

\noindent Thus only when we have $v^{rh} =t$ can we successfully add element $e$ into the payload $E$. Otherwise, it is to be wiped off by some $rmv$ operation and can be safely ignored.

\subsection{From \rwf-Set to \rwf-Skeleton} \label{SubSec: RWF-Skeleton}

The \rwf-Set can be augmented to store application-specific values. Since the conflict concerning the existence of elements is handled by the \rwf-Set, the user can focus on the conflicts concerning the value of elements.

The specification of our \rwf-Set is $\{e|\exists add(e):\forall rmv(e):rmv(e)\vis add(e)\}$. This is different from the specification of the existing Remove-Win Set\cite{Zaw15}, which is $\{e|\exists add(e)\wedge \forall rmv(e):\exists add(e): rmv(e)\vis add(e)\}$.
The existing remove-win strategy actually records all the newest add/remove operations and decide whether the element exist afterwards, which is mostly like the add-win strategy of OR-Set\cite{Shapiro11b} with different concurrent add/remove preference. This kind of strategies that record operations and decide afterwards is not suitable for handling the value of elements, which is needed to further augment the set into container CRDT design framework. Because the validity of value depends on the existence of the element, which can not be decided until all relevant add/remove operations are recorded. This increases the complexity of designing the container type CRDT. In our \rwf-Set, system execution is segmented into phases by more powerful remove operations. This helps designing the \rwff for container type CRDTs.

The conflict resolution concerning values can be destructed into three basic cases. Thus the \rwf-Skeleton is proposed, where three open terms are left for the user to develop stubs containing their own conflict resolution logics, as shown in Algorithm \ref{A: RWF-Skeleton}. With the \rwf-Skeleton, the concrete design of an \rwf-DT can be obtained by specifying how the values are initialized and updated via the \rwf-DT APIs and plugging the conflict-resolution stubs.

We first briefly overview conflict resolution involving remove operations. Then we focus on the three basic cases of conflict resolution among non-remove operations. An exemplar \rwf-RPQ design is presented here, and its implementation is presented in Section \ref{Sec: Impl}. More exemplar designs are presented in Appendix A-D in \cite{RWF-TR20}.

\begin{algorithm}[htbp]
    \DontPrintSemicolon
    \caption{\rwf-Skeleton}
    \label{A: RWF-Skeleton}
    \Payload $E$: set of $(e,\textcolor{\frcolor}{p_{ini}})$ tuples, $T$: set of $(e,t)$ tuples, $V$: set of $(id, \textcolor{\frcolor}{v_{inn}},\textcolor{\frcolor}{v_{acq}})$ 	tuples \;
    \Initial $E = \emptyset, T = \emptyset, V = \emptyset$ \; 
    \Update({$ add(e) $})
    {
        \AtSource({$(e)$})
        {
            \Pre $e$ is not in the data collection\;
            \Let $v^{rh} = t$ s.t. $(e,t) \in T$ \tcp*{$v^{rh} = \vec{0}$ if there is no $(e,t)$ in $T$.}
            \Let $p_{ini}$ be id of the initiator of this operation \;
        }
        \DownStream({$(e, p_{ini}, v^{rh})$})
        {
            $rmv(e, v^{rh})$ \tcp*{Execute the \textbf{effect} part of $rmv(e)$ using $v^{rh}$.}
            \Let $t : (e, t )\in T$ \tcp*{$t = \vec{0}$ if there is no $(e,t)$ in $T$.}
            \uIf(\tcp*[f]{The remote replica and the $add$ operation are in the same phase.}){$v^{rh} = t$}
            {
                $E := E \cup \set{(e, p_{ini})}$ \; 
                \frm{determine the innate value $v_{ini}$ for $e$} \tcp*{Resolve possible conflicts between concurrent $add$s, using $p_{ini}$ to obtain the replica information.}
            }
        }
    }
    \Update({$rmv(e)$})
    {
        \AtSource({$(e)$})
        {
            \Pre $e$ is in the data collection\;
            \Let $v^{rh} = t$ s.t. $(e,t) \in T$ \tcp*{$v^{rh} = \vec{0}$ if there is no $(e,t)$ in $T$.}
            \Let $p_{ini}$ be id of the initiator of this operation \;
            $v^{rh}[p_{ini}] := v^{rh}[p_{ini}]+1$\;
        }
        \DownStream({$(e, v^{rh})$})
        {
            \Let $t:(e,t)\in T$\tcp*{$t = \vec{0}$ if there is no $(e,t)$ in $T$.}
            \uIf(\tcp*[f]{There are unrecorded $rmv$ operations in $v^{rh}$.}){$\exists k: t[k] < v^{rh}[k]$}
            {
                Remove $(e,p_{ini})$ from $E$ if any\; 
                Remove $(e, v_{inn}, v_{acq})$ from $V$ if any\; 
                \Let $t': \forall k:t'[k]:=\max(v^{rh}[k], t[k])$\;
                $T:=T\setminus\{(e,t)\}\cup\{(e,t')\}$\;
            }
            
        }
    }
    \Update({$upd(e)$})
    {
        \AtSource({$(e)$})
        {
            \Pre $e$ is in the data collection\;
            \Let $v^{rh} = t$ s.t. $(e,t) \in T$ \tcp*{$v^{rh} = \vec{0}$ if there is no $(e,t)$ in $T$.}
        }
        \DownStream({$(e, v^{rh})$})
        {
            $rmv(e, v^{rh})$  \tcp*{Execute the \textbf{effect} part of $rmv(e)$ using $v^{rh}$.}
            \Let $t : (e, t )\in T$ \tcp*{$t = \vec{0}$ if there is no $(e,t)$ in $T$.}
            \uIf(\tcp*[f]{The remote replica and the $add$ operation are in the same phase.}){$v^{rh} = t$}
            {
                \frm{Modify the acquired value $v_{acq}$ for $e$} \tcp*{Resolve possible conflicts between concurrent $upd$s, using $p_{ini}$ to obtain the replica information if necessary.}
            }
        }
    }
\end{algorithm}

\subsubsection{Remove-Win Resolution}

The \rwf-Skeleton has the new value-updating operation $upd$, which enables the user to modify the values of existing data elements. Comparing with the \rwf-Set, the $add$ operation in the \rwf-Skeleton not only creates a data element, but also sets its initial value. Owing to the remove-win strategy, the conflict resolution between remove and non-remove operations ($add$ and $upd$) are principally the same. The $rmv$ operations win, and the effects of (concurrent or causally visible) non-remove operations are wiped off. 

The execution is still segmented into phases by $rmv$ operations. When executed on a remote replica, each non-remove operation carries the rh-vec, uses the vector to firstly execute the missing $rmv$ operations at the \textsf{effect} part of this operation and then takes effect only if this operation is in the same phase with the replica.

\subsubsection{User-specified Resolution}

With the help from the \rwf-Set, the user only needs to care about the conflicts concerning data values among non-remove operations within each phase. Two types of non-remove operations, $add$ and $upd$, may modify the value and potentially cause conflicts. Thus, there are three different types of possible conflicts to be considered, as detailed one by one below.


\noindent \textit{Add-add resolution}. 
When two different $add$ operations both add the same element, but setting different initial values, there will be a conflict. An open term is left in the skeleton (Line 13 in Algorithm \ref{A: RWF-Skeleton}) to let the user specify how to handle this conflict. 
Principally, the user must use certain information of the initiating replicas, in order to differentiate concurrent $add$ operations. Thus, the payload $E$ not only contains the element ID, but also contains $p_{ini}$, the ID of the initiating replica. The $p_{ini}$ can be thought as a handler, with which the $add$ operation can access any information of the replica necessary to differentiate concurrent $add$ operations. For example, the user may specify ``larger replica ID wins", assuming that the replica IDs are totally ordered. Thus the initial value of element is set to the value from the $add$ operation initiated by the replica with larger ID.


\noindent \textit{Upd-upd resolution}.
The value of elements may be modified by application-specific $upd$ operations. Conflict between $upd$ operations is to be resolved by user-specified resolution logic (Line 35 in Algorithm \ref{A: RWF-Skeleton}). For example, for a list, the user may employ an \textit{operational transformation} algorithm to decide the results of all possible conflicting list updates ($insert$ and $delete$) \cite{Attiya16, Wei18}. As for a priority queue, the value increase/decrease operations naturally commute. Thus no resolution is needed, as detailed in Appendix A of \cite{RWF-TR20}.


\noindent \textit{Add-upd resolution}.
Though the $add$ operation and the $upd$ operation both can modify the value of data items, they have different types of user intention behind them. Specifically, the $add$ operation initializes the value. It has semantics similar to those of value assignments. The $upd$ operation modifies value. The semantics is application-specific, and usually are different from those of value assignments. For example, priority values of elements in a priority queue are often modified by increase or decrease of the (numerical) priority values.

According to the two (often) different types of user intentions, we divide the value of an element into the \textit{innate value} and the \textit{acquired value} (payload $V = (id, v_{inn}, v_{acq})$ in Line 1 in Algorithm \ref{A: RWF-Skeleton}). Accordingly, the innate value stores the initial value of the element brought by $add$ operations, whose conflict have been resolved. And the acquired value stores the relative change of the actual value of the element from the innate value brought by $upd$ operations. The result of \textit{upd-upd resolution} related to the value of the element is stored here.

Thus, the conflict between an $add$ and an $upd$ operation is resolved by dividing the data value into two parts, one part for each operation. 
And the actual value of the element is the summary of the innate value (initial value set when added) and the acquired value (relative change that summarizes all $upd$ operations).
Such division of value is rather conceptual here, and requires further implementation by the CRDT designer.

%
%

\section{\rwf-DT Implementation} \label{Sec: Impl}

In this section, we explain how to use the \rwff design framework in practice, with an exemplar priority queue implementation over Redis. More details of another list implementation can be found in Appendix C of \cite{RWF-TR20}. Redis is a widely-used in-memory data type store. It adopts the master-slave architecture\footnote{\scriptsize The enterprise version of Redis supports the multi-master architecture, and uses CRDT to handle conflicts. However, this version is not open source.}. We modify Redis to work in the multi-master mode, and CRDTs are used for conflict resolution. Note that the adoption of \rwff is orthogonal to that of the underlying data store, and \rwff can be applied to other data type stores like Riak \cite{Riak19}. All the implementation can be found at the GitHub repository \cite{CRDT-Redis}.

The implementation of an \rwf-DT has the ``onion" structure, and proceeds through three levels -- the CRDT level, the \rwff level and the DT level, as shown in Fig. \ref{F: 3Level-Framework}. In the outermost level, the data type is first a CRDT. The basic template for local processing and asynchronous propagation of data updates is specified. In the middle level, the data type uses \rwff for conflict resolution. Common metadata and conflict resolution logics following the \rwf-Skeleton are specified. In the innermost level, definition of the specific data type and user-specified logics for conflict resolution are provided. In this section, we introduce these three levels one by one.

\subsection{CRDT Level Implementation}

\begin{figure}[htbp]
    \centering
    \includegraphics[width=\linewidth]{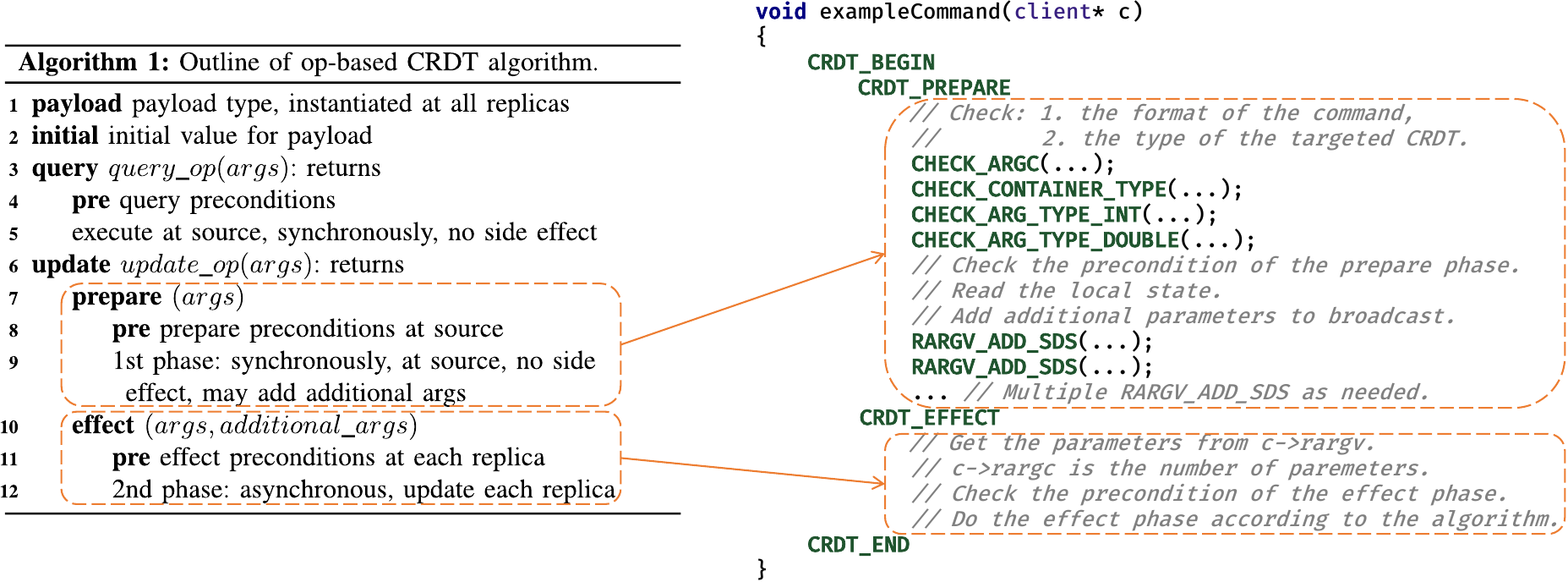}
    \caption{Implement one CRDT operation with framework.}
    \label{F: CRDT-Level}
\end{figure}

In the outermost level, we implement the CRDT framework as a code template over Redis, as shown in Fig. \ref{F: CRDT-Level}. Operations which are common to different CRDT designs are abstracted as four macros and 2 types of tool functions, as detailed below.  

\subsubsection{\texttt{CRDT\_BEGIN}}

The \texttt{CRDT\_BEGIN} macro checks if the data store (Redis instance) works in the multi-master replication mode. If not, it is invalid to use CRDTs.

\subsubsection{\texttt{CRDT\_PREPARE}}

When receiving a request, the server first needs to check its type, i.e., a \textit{client request}, or a \textit{server request}. In case of a client request, the server proceeds to the \textsf{prepare} part processing. For a server request, the server directly jumps to the \textsf{effect} part. In the local processing (in the \textsf{prepare} part) of a client request, two types of operations are common to different CRDTs. 

First, the server needs to check whether the client is using the correct API the server provides. In case the API is correct, the server further checks whether the client is providing correct parameters for the API invocation. Note that the number of parameters and the type of each parameter can only be decided in the DT level. Now in the CRDT level, we provide tool functions, which encapsulate the logic for checking the number of parameters, while the actual number of parameters to be checked will be passed in as parameters later in the DT level. We also provide tool functions for parameter type checking, for widely used types such as \texttt{INT} and \texttt{DOUBLE}. The user just chooses the correct tool function and passes the correct parameter in the DT level. In case the parameter type checking function is not provided, e.g. checking functions for user defined types, the user needs to implement the checking functions themselves, following the existing tool functions.

Second, the local processing needs to prepare multiple parameters to be broadcast for the remote processing. A dynamic array is used to contain any number of parameters, and in our Redis implementation, each parameter is in the Simple Dynamic String (SDS) format defined by Redis. For any type of parameters to be broadcasted, the user only needs to provide the serialization and de-serialization functions to and from the SDS format.


\subsubsection{\texttt{CRDT\_EFFECT}}

In the \textsf{effect} part, the server first acknowledges its reception of the server request. The concrete logics for the processing, mainly the conflict resolution logics, are filled in later in the \rwff level and the DT level.

\subsubsection{\texttt{CRDT\_END}}

At the end of a CRDT operation, the server acknowledges the client or server request.

\subsection{\rwff Level Implementation -- Data Element Definition}

In this section, we discuss the data element definition in the \rwff level, which extracts the common characteristics of the data container type we focus on. We first discuss the innate and acquired values of concrete data. Then we discuss the metadata for the remove-win conflict resolution.

\subsubsection{Innate and Acquired Values}

Each \rwf-DT shares the common nature of being a container of data elements. Each data element has its ID, which identifies the existence of the element. How the ID is defined, e.g. using a 64 bit string or a long integer, will be decided in the DT Level. 

Each element in the container has its value. The value is initialized by $add$ operations, and is then updated by the $upd$ operations (together with the conflict resolution logics). However, one important common pattern is that the value of each data element has two different types of constituents, with different intentions behind (see detailed discussions in Section \ref{SubSec: RWF-Skeleton}). One is innate value. It is often associated with initialization. The intention behind the initialization is value assignment. The new initialization should overwrite the old one. However, the concrete definition of `old' and `new' is user-specified since there may be concurrent initializations (later in the DT level). The other one is acquired value. The conflict resolution logic could be arbitrary and user-defined. However, it is often different from the conflict resolution logic for innate values. The classification of innate and acquired values further simplifies the development of conflict resolution logics. 

\subsubsection{Metadata for Conflict Resolution}

The conflict resolution is based on the pre-defined remove-win strategy. Thus each element has $pid$ and $current$. The $pid$ is the ID of the replica which accepts the request from the client. This $pid$ info identifies each replica. This info is leveraged to break the symmetry between concurrent (conflicting) updates.

\begin{figure}[h]
    \centering
    \includegraphics[width=\linewidth]{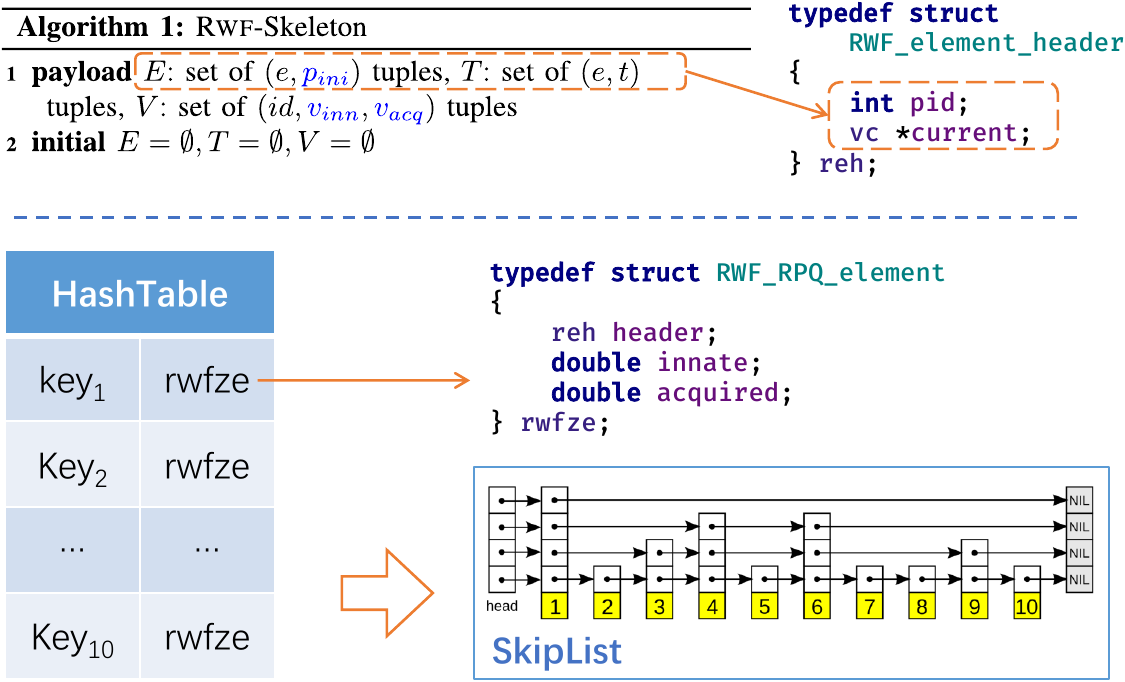}
    \caption{The data storage implementation of \rwf-RPQ.}
    \label{F: Data-Def}
\end{figure}

The $current$ is the rh-vec timestamp. As in the \rwf-Skeleton, the rh-vec is encoding of the remove history, which is essential to the conflict resolution following \rwf. We define the struct \texttt{\rwf\_element\_header} containing this metadata, as shown in Fig. \ref{F: Data-Def}. All \rwf-DT metadata structs extend the header to contain specific (innate and acquired) values.

\subsubsection{Data Organization on the Server Replica}

Here we use a hash map to store the metadata of a data type following \rwf. This hash map can be used to get the element in the container by its key. Other data structures can be used for the \rwf-DT if needed. For example, our exemplar \rwf-RPQ implementation additionally uses a skiplist \cite{Pugh90} to maintain the order of elements.

\subsection{\rwff Level Implementation -- Conflict Resolution}

\begin{figure}[htbp]
    \centering
    \includegraphics[width=\linewidth]{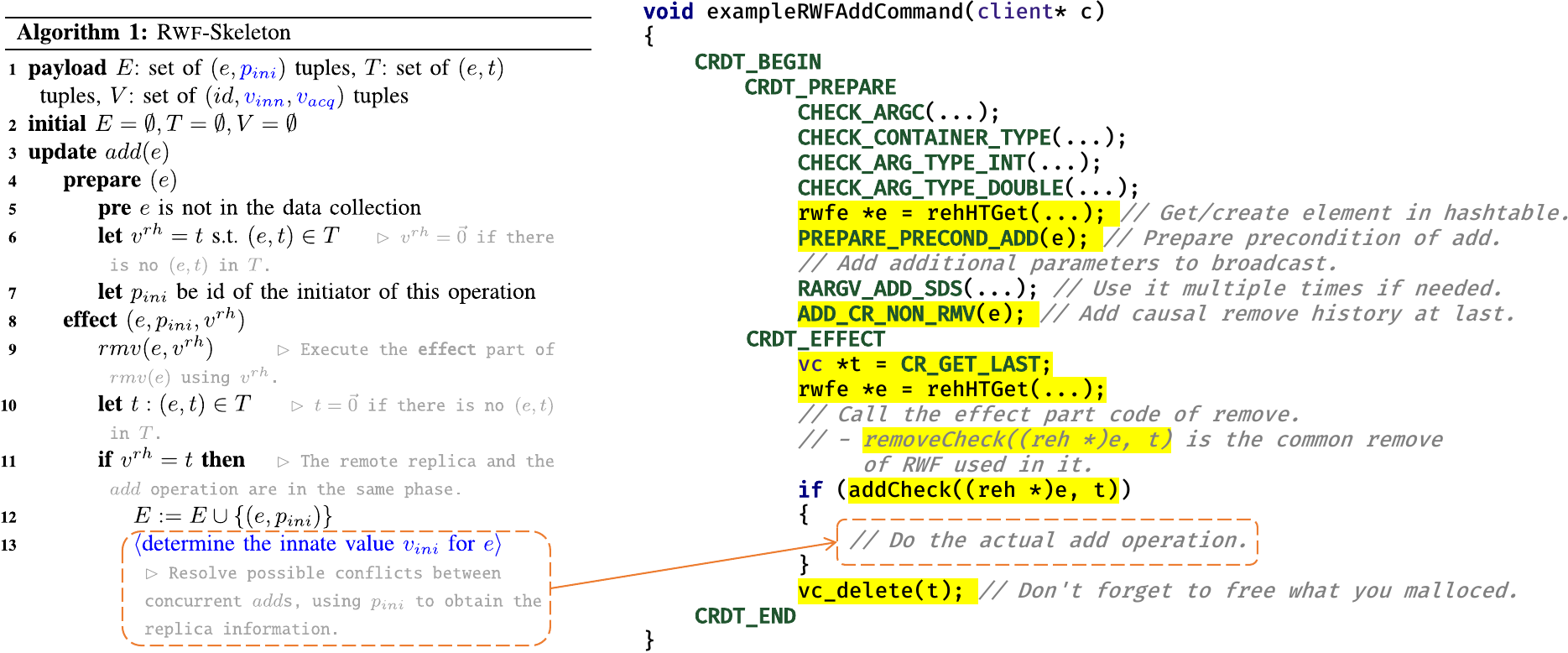}
    \caption{Implement the add operation of a \rwff CRDT using the framework.}
    \label{F: RWF-Level}
\end{figure}

In the \rwff level, the CRDT template (in Fig. \ref{F: CRDT-Level}) is further extended to include the data definitions and conflict resolution operations which are pertinent to the remove-win resolution strategy, as highlighted in Fig. \ref{F: RWF-Level}. Here we use the $add$ command as an example to illustrate the \rwff level implementation.

\subsubsection{Prepare}

In the local processing of a client request, we first need to get the element in the hash table. Though the specific data element type may vary, getting the handler of one data element in the hash table has the generic pattern. Specifically, we first get the correct data container in the data store (we may have multiple data containers working in the data store). We then get the element by its key. We also need to get the handler of the local data structure for maintaining the structure among data elements. In the \rwff level, we provide tool function \texttt{rehHTGet($\cdots$)}\footnote{\scriptsize See detailed comments of the ``\texttt{rehHTGet}" function in ``redis-6.0.5/src/RWFramework.h" at the repository \cite{CRDT-Redis}.}, and the user further provides parameters as required in the DT level.

Before doing the actual processing, we need to first guarantee that certain precondition holds. In the \rwff level, we implement two common precondition checking functions widely used in data container types. Specifically, data container operations often need to ensure that the current element is or is not in the container. We implement two tool functions for these two types of checking. Other user-defined precondition checking can be supplemented by the user in the DT level.

In the end of the local processing, the remove history of data element needs to be updated, which is essential to the remove-win conflict resolution. The tool function for updating the remove history is implemented in the \rwff level.

\subsubsection{Effect}

To conduct remove-win conflict resolution, the replica should first get the remove history (rh-vec) of the element under processing. The tool functions/macros of getting and deleting the rh-vec is provided in the \rwff level. Given the rh-vec of the remote replica, the current replica needs to get the element from the hash table. This is principally the same with the \texttt{rehHTGet($\cdots$)} operation in the \textsf{prepare} part.

As discussed in Section \ref{SubSubSec: Conf-Res}, to cope with the late arrival of messages, the replica should check the remove history and do the missing remove operation first. Note that the remove operation not only eliminates the current element. It also needs to update the data structure after the delete operation. This update is provided in the DT level. 

Before doing the actual processing, the replica needs to check whether the remove operation and the current replica are in the same phase, by comparing the rh-vecs (see Line 11 of Algorithm \ref{A: RWF-Skeleton}). After the checking, the actual processing can be conducted. In our example, we provide the ``\texttt{addCheck}" function. Similarly for $rmv$ and $upd$ operations, we provide the corresponding ``\texttt{rmvCheck}" and ``\texttt{updCheck}".

\subsection{DT Level Implementation -- an RPQ Example}

Here we give an example of how to implement a replicated priority queue, denoted as \rwf-RPQ, using the \rwf.
We first ``inherit" the \texttt{\rwf\_element\_header} to define the metadata struct of elements \texttt{rwfze}, as shown in Fig. \ref{F: Data-Def}. As an \rwf-RPQ element, it further contains the innate and acquired values. 
Each data element has its ID and value (defined in \texttt{rwfze}). The key-value pairs (ID, \texttt{rwfze}) are stored in the hash table.
For the priority queue, each server uses the skip list to organize the elements with their priorities. Local organization of data elements is orthogonal to the design of the \rwf-DT.


The users provide parameters to the tool functions. They may also implement the concrete ``\texttt{removeFunc}" for deleting an element from a data structure. Finally the users provide the logics for conflict resolution.

The development task is greatly simplified. The user only needs to adopt the template, choose the tool functions, and provide parameters to the functions. The user-defined logics are then supplemented in the indicated places.

\begin{figure}[h]
    \centering
    \includegraphics[width=\linewidth]{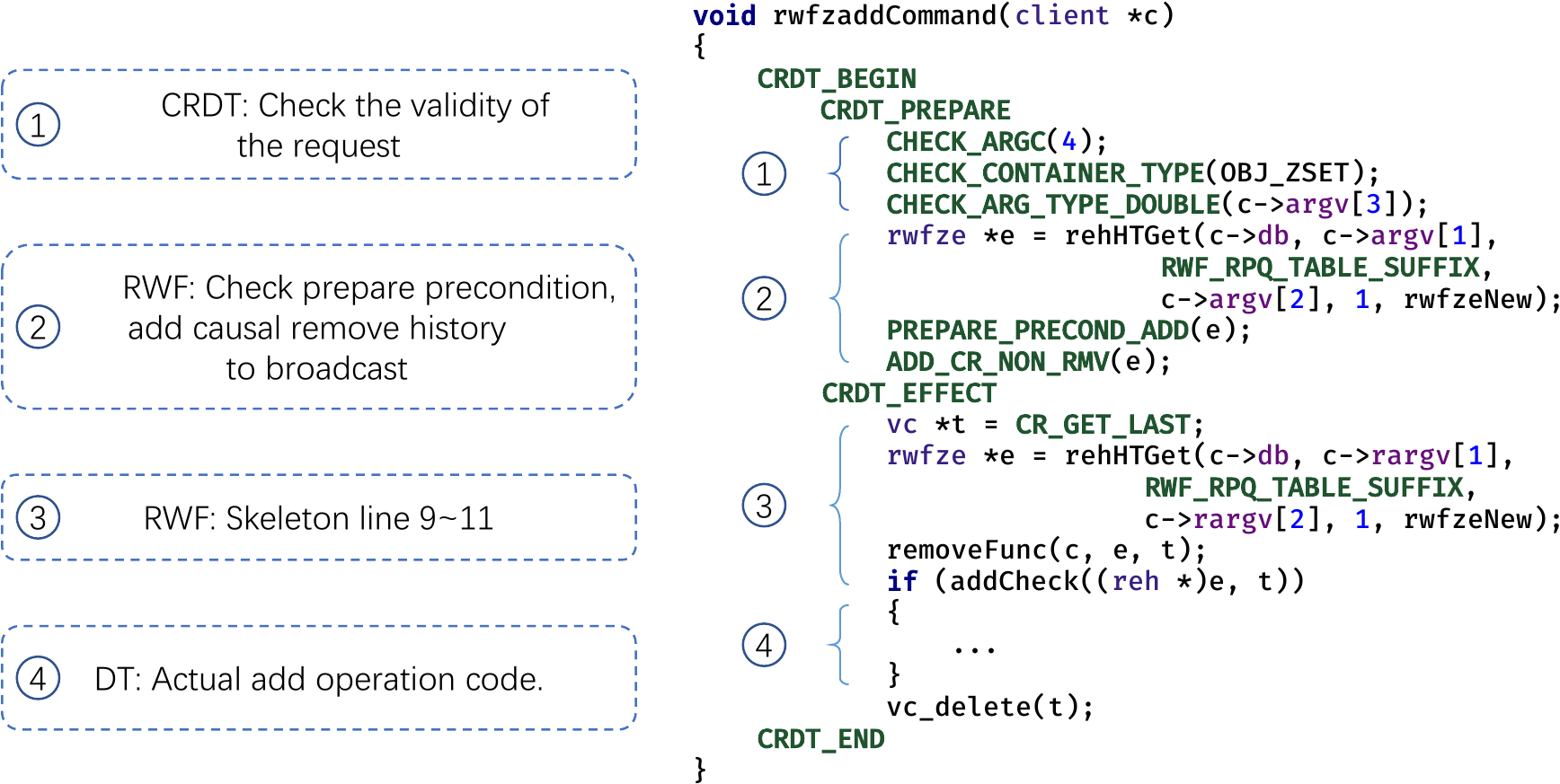}
    \caption{The implementation of add operation of \rwf-RPQ in Redis.}
    \label{F: DT-Level}
\end{figure}

\section{Experimental Evaluation} \label{Sec: Exp}

In this section, we first present the experiment setup and design. Then we discuss the evaluation results.

\subsection{Experiment Setup}

\begin{figure}[ht]
    \centering
    \includegraphics[width=.9\linewidth]{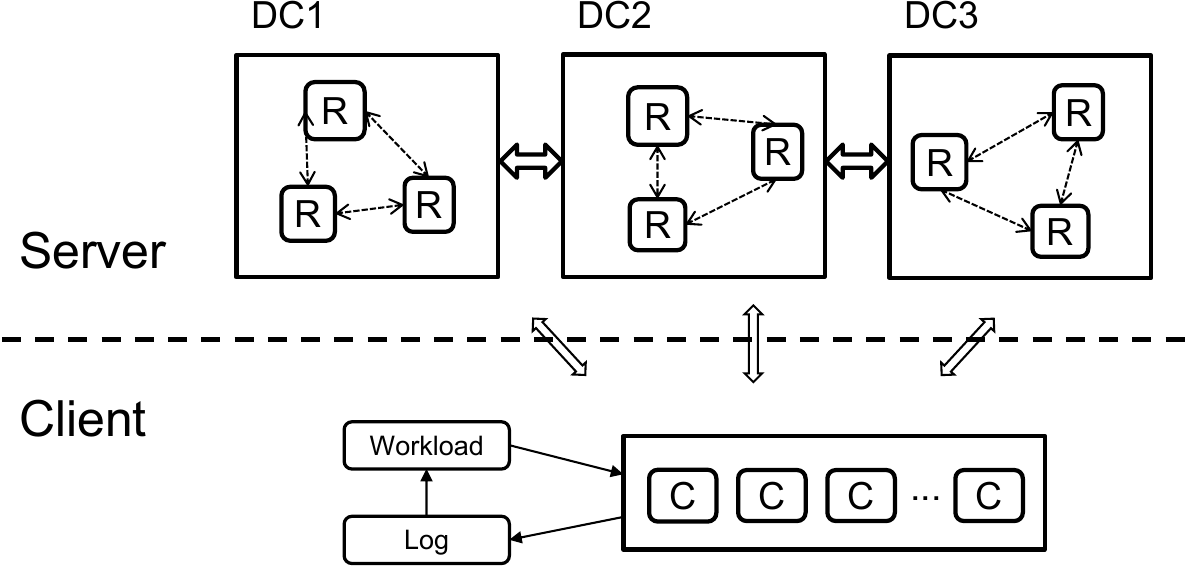}
    \caption{Experiment setup.}
    \label{F:Exp-Arch}
\end{figure}

The experiment is conducted on a workstation with an Intel i9-9900X CPU (3.50GHz), with 10 cores and 20 threads, and 32GB RAM, running Ubuntu Desktop 16.04.6 LTS.
We run all server nodes and client nodes on the workstation. Logically we divide the Redis servers into 3 data centers as shown in Fig. \ref{F:Exp-Arch}. Each data center has 3 instances of Redis.
We use traffic control (TC) \cite{TC20} to control the network delay among Redis instances. The default inter-data center communication delay follows \nordis{50}{10}\footnote{\scriptsize \nordis{\mu}{\sigma} stands for the normal distribution, where $\mu$ is the mean and $\sigma$ is the standard deviation.}, while the default intra-data center delay follows \nordis{10}{2} (the time unit is \textit{ms}). We use this set of network delay based on our experience.

The clients obtain when and what operations to issue to the servers from the workload module. This module generates workloads of different patterns. The clients record statics about how operations are served by the servers in the log module. When generating the operations, the workload module needs to query the log module, to obtain current status of the CRDT. This is because the workload module may need to intentionally generate conflicting update operations. Also, it needs to prevent invalid operations such as removing an element that does not exist in the CRDT.

\subsection{Experiment Design}

We design replicated priority queue and replicated list, using both the existing remove-win strategy \cite{Zaw15} and our \rwff design framework (namely the Remove-Win RPQ, the \rwf-RPQ, the Remove-Win List and the \rwf-List). The design and implementation of the data types used in the experiments are all available online\footnote{\scriptsize See detailed discussions on the design in Appendix A-D of \cite{RWF-TR20}. The source codes are also available in the repository \cite{CRDT-Redis}.}.

The key space for elements in the RPQ has size 200,000. The workload module randomly chooses elements to be added from all possible ones. The $inc$ and $rmv$ operations are conducted on random elements in the RPQ. The initial values of elements are randomly chosen from integers ranging from 0 to 100. The value increased is randomly chosen from -50 to 50.

Because the key space of RPQ in our experiment is relatively large, the probability of generating conflicting operation pairs containing $add$ on the same element is low, as we randomly choose elements from the key space for $add$. We intentionally create such conflict operation pairs to evaluate the performance of an RPQ. When the workload module generates the latest operation $o$, it will pair $o$ with all operations which are less than $\mu$ units of time before $o$. Here, $\mu$ is the average message delay of intra-data center communication. The workload module is concerned of $add$-$add$ and $add$-$rmv$ pairs. All such pairs have probability 15\% to execute on the same data element. Note that we do not explicitly control the conflict for $inc$-$rmv$ pairs. It is because there will be fairly high probability of such conflicts, as they are conducted only on the elements that are already in the RPQ. All workloads we consider have 59\%--89\% operations which are $inc$ or $rmv$. 

The replicated lists are targeted at strings of text chars in collaborative editing scenarios. We use $(clientID,num)$ pairs as the keys of the elements in lists. We generate a new key for each $add$ operation, and
all \textit{undo} and \textit{redo} operations are translated into $add$ and $rmv$ operations. 
To exercise the conflict resolution strategies, 50\% $add$ operations will add previously removed elements, and the rest of $add$ operations will add new elements. 
There are 6 properties for elements in the list: font(0-9), size(0-99), color(24 bits), bold(Y/N), italic(Y/N), underline(Y/N). The $upd$ operation randomly chooses one property to update. Both the initial properties and the $upd$ operation parameters are chosen at random. The $upd$ and $rmv$ operations are conducted on random elements which are currently in the list. We do not need to intentionally create conflicting operations for lists, as the probability of conflict is fairly high.

Since the CRDTs serve operations instantly by design, they have statistically the same performance in terms of query / update delay. However, there is the intrinsic tradeoff between data consistency and response latency. Thus we need to measure the data consistency, in order to show how much data consistency is sacrificed to obtain the performance in response delay. As for the priority queue, we measure the difference between the return value of $get\_max$ and the real $max$ value. The read-time order in which queries/updates are logged on the client side is approximately the order they are served by the servers. We use this total real-time order to decide the status of the priority queue and calculate the correct $max$ values. As for the list, we also use the real-time order on the client side to obtain the linearized list. We measure the edit distance between the list on the server and the list linearized on the client side. We further measure the edit distance between lists from different servers. Also we record the metadata overhead for resolving conflicts by the CRDTs under evaluation. The metadata overhead is averaged among all elements in the data container.

We use two types of workload patterns for both RPQs and Lists. First, we have the \textit{add-rmv dominant} pattern where 41\% operations are $add$, 39\% operations are $rmv$ and 20\% operations are $upd$. Second, we have the \textit{upd dominant} pattern where 80\% operations are $update$, 11\% operations are $add$ and 9\% operations are $rmv$\footnote{\scriptsize We make the $add$ operations appear slightly more than $rmv$ to prevent the RPQ from being often empty.}. We generate 4,000,000 operations in total for RPQs, 10,000 operations per second. As for lists, the number of operations generated is 400,000, 1000 operations per second.

\subsection{Evaluation Results}

We list the average performance in terms of data inconsistency of all data types in Table \ref{T: Consis}.

Then we discuss the evaluation results for the priority queues and lists in detail.
Please note that, more evaluation results and the corresponding discussions are provided in Appendix E of \cite{RWF-TR20}, due to the limit of space.

\begin{table}[tbp]
    \centering
    \caption{Data inconsistency on average. `\textit{r}' means Remove-Win CRDT, and `\textit{rwf}' means \rwf-DT. `\textit{$upd$-dom}' stands for the $upd$-dominant pattern, and `\textit{$a/r$-dom}' stands for the $add$/$rmv$-dominant pattern.} 
    \label{T: Consis}
    \begin{tabu}{X[1.9,m,l]*2{X[0.8,m,l]}*2{X[1.2,m,l]}*2{X[1.35,m,l]}}
        \toprule      
        & \multicolumn{2}{l}{RPQ {\scriptsize (Fig.\ref{F:exp-rpq})}}& \multicolumn{2}{l}{List-local {\scriptsize (Fig.\ref{F:exp-list})}} & \multicolumn{2}{l}{List-replica {\scriptsize (Fig.\ref{F:exp-list-cmp})}} \\
        \cmidrule(r){2-3}\cmidrule{4-5}\cmidrule(l){6-7}
        & r & rwf  & r & rwf & r & rwf \\
        \midrule
        {\small $upd$-dom} & 14.8 & 4.0 & 449.8 & 301.0 & 7.8 & 7.5   \\
        {\small $a/r$-dom} & 31.4 & 38.4 & 15416.1 & 11725.2 & 12.4 & 14.8 \\
        \bottomrule    
    \end{tabu}
\end{table}

\subsubsection{Replicated Priority Queue}

\begin{figure}[h]
    \centering
    \includegraphics[width=\linewidth]{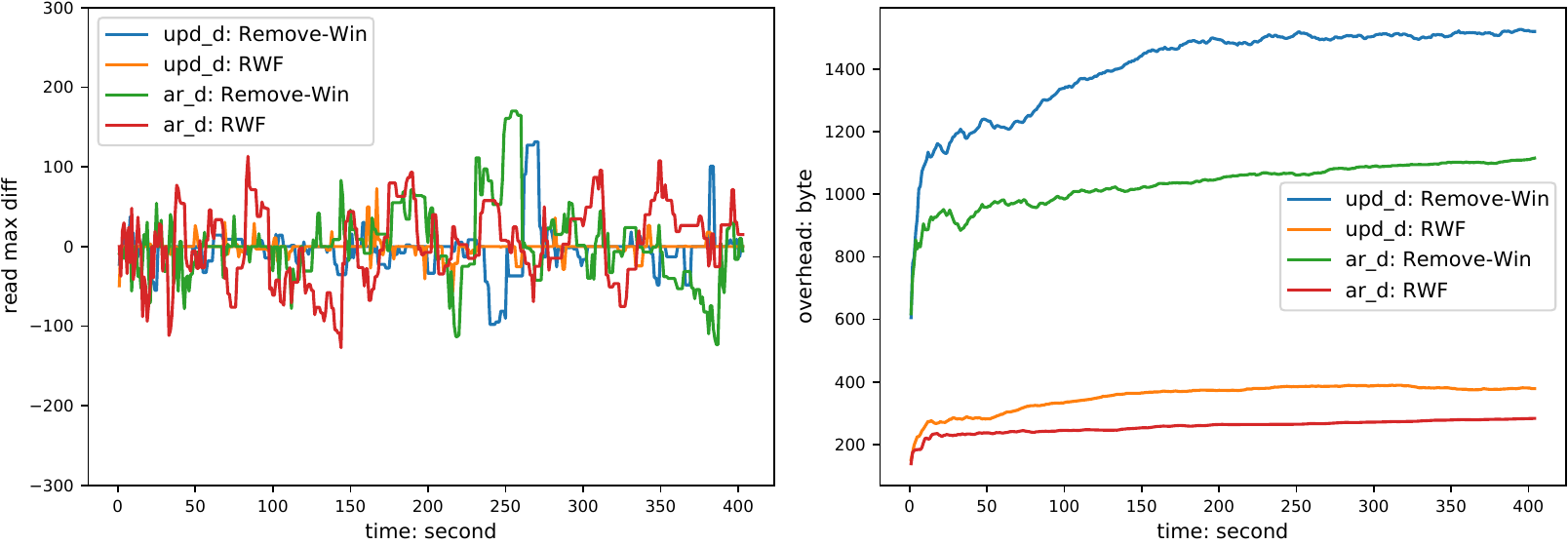}
    \caption{The performance of RPQs, comparing max value read from the server with the max value of local linearized queue.}
    \label{F:exp-rpq}
\end{figure}

We first compare the return value of $get\_max$ from server, and the max value of the centrally linearized queue. As shown in Fig. \ref{F:exp-rpq}, the difference vibrates mostly between -100 and 100. This is relatively small, considering the increase value we generate are chosen randomly between -50 and 50. According to evaluation results in Fig. \ref{F:exp-rpq} and Table \ref{T: Consis}, two RPQs act similarly considering the read max difference. The $add$/$rmv$-dominant workload pattern causes more differences. This is mainly because, in the $add$/$rmv$-dominant workload, data items enter and leave the queue more frequently, while in the $upd$-dominant workload, data elements in the queue are relatively stable, only their priority values change more frequently. Thus in the $add$/$rmv$-dominant workload, the max priority value in the queue are frequently changed abruptly, due to the add and deletion of data elements\footnote{\scriptsize We also compare the difference between two queues. The results are principally the same with those by comparing the replicated queue and the linearized queue. The results are shown in Appendix E in \cite{RWF-TR20}.}.

As for the metadata overhead of two RPQs, it slowly increases as more operations are executed. We do not have garbage collection for the removed elements, thus needing to store their tombstones. Such removed elements require more storage as more $rmv$ are executed.
The metadata overhead is higher in the $add$/$rmv$-dominant pattern, because the RPQ needs to store more conflict resolution data for $add$/$rmv$ operations than for $inc$ operations.
The \rwf-RPQ has less metadata overhead than the Remove-Win RPQ, mainly because the latter needs more space to guarantee the causal delivery of messages.

\subsubsection{Replicated List}

\begin{figure}[h]
    \centering
    \includegraphics[width=\linewidth]{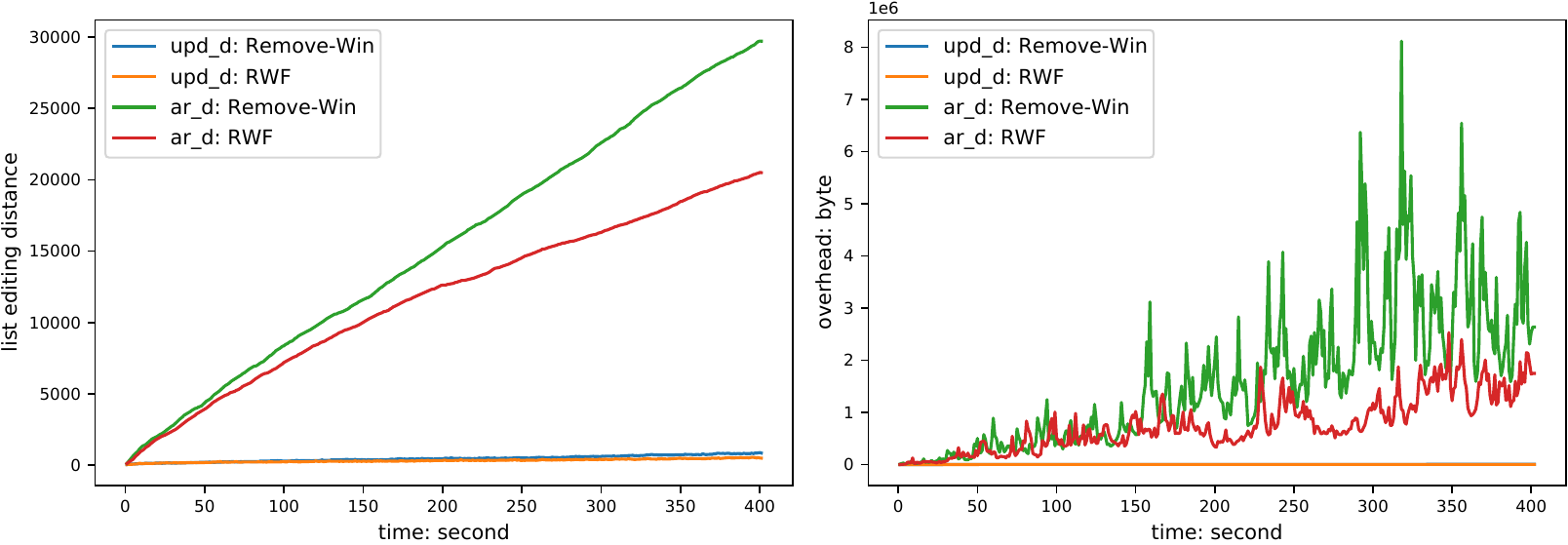}
    \caption{The performance of lists, comparing the edit distance between the list read from the server and the local linearized list.}
    \label{F:exp-list}
\end{figure}

We first compare the list on the replicas with the list linearized on the client side. The results are shown in Fig. \ref{F:exp-list}. The edit distance increases as more operations are executed. This is because the CRDTs only guarantee eventual convergence. The replica is not guaranteed to be the same with (or similar to) the linearized one. The edit distance of the $upd$-dominant pattern is relatively small. This is because $upd$ operation does not affect the order of elements. Less $add$/$rmv$ operations mean that the server will execute $add$/$rmv$ in a more sequential manner, and need less conflict resolution. 

\begin{figure}[h]
    \centering
    \includegraphics[width=\linewidth]{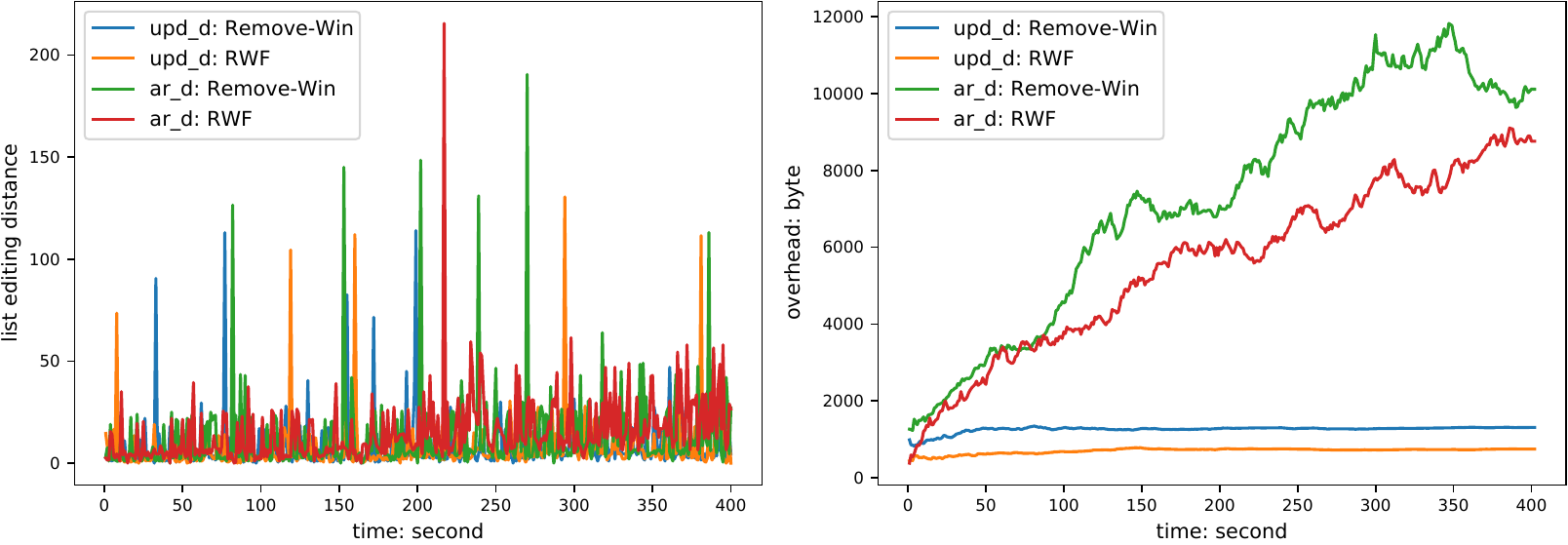}
    \caption{The performance of lists, comparing the edit distance between two lists read from different servers at the same time.}
    \label{F:exp-list-cmp}
\end{figure}

We then compare the lists on different servers at the same time instant. As shown in Fig. \ref{F:exp-list-cmp} and Table \ref{T: Consis}, both Remove-Win List and \rwf-List perform well. The distances of two lists are mostly within 50, and two lists quickly converge. The distance of the $upd$-dominant pattern is slightly small, as shown in Table \ref{T: Consis}. This is also because less $add$/$rmv$ operations induce less divergence between the replicas.

As for the metadata cost, the overhead slowly increases as we need to store the tombstone of the removed elements. The overhead is much lower in the experiment of comparison between replicas (Fig \ref{F:exp-list-cmp}), because here we make 50\% $add$ to add previously removed elements, causing their tombstones to be efficiently reused. 
The metadata overhead is much lower in the $upd$-dominant pattern. 
Similar to the RPQ case, conflict resolution data needed for $upd$ is much less for that of $add$/$rmv$ operations.
Moreover, the Remove-Win List needs to maintain causal message delivery, which causes higher metadata overhead.

\section{Related Work} \label{Sec: RW}

Conflict resolution is the essential issue in the design of CRDTs. For data container types, the dual add-win and remove-win strategies are intuitive and widely used. The Add-Win Set proposed in \cite{Shapiro11b} lets each $rmv$ operation record all $add$ operations it has seen. The effect of a $rmv$ operation is limited to the $add$ operations it has seen, which makes the $add$ operation win over the concurrent $rmv$. The design of the Remove-Win Set proposed in \cite{Zaw15} is dual to that of the Add-Win Set. Each $add$ operation is required to record all the $rmv$ operations it has seen. The effect of $add$ operations is limited to these $rmv$ operations it has seen, which makes the $rmv$ operation win over the concurrent $add$. In existing add-win and remove-win sets, all operations are recorded in the execution and a total order among all operations is derived to interpret the state of each replica. In our \rwff design framework, non-remove operations which are concurrent with a remove operation are pruned from the execution under concern. Thus no conflict will occur concerning remove operations. The remove-win strategy used in \rwff further utilizes the potential of the remove-win strategy, thus better supporting a design framework. Experiments show that the semantics of \rwf-DTs are statistically similar to CRDTs using the existing remove-win strategy.

Existing CRDT designs are often obtained via derivations from seminal and widely-used designs, which motivates us to propose our design framework. In the area of collaborative editing, the WOOT model is proposed, which essentially designs a conflict-free replicated list \cite{Oster06}. 
Multiple improved designs following WOOT were proposed, including WOOTO
and WOOTH \cite{Ahmed11}.
In the area of computational CRDTs, for a class of CRDTs whose state is the result of a computation over the executed updates, a brief study is presented in \cite{Navalho15} and three generic designs are proposed. The non-uniform replication model is further proposed to reduce the cost for unnecessary data replication, which is often seen in computational scenarios \cite{Cabrita17}. Though existing derivations of CRDT designs are mainly driven by  the application scenarios, our \rwff design framework focuses on the data type itself. \rwff focuses on the widely-used data collection type and can be used in a variety of application scenarios.


\section{Conclusion} \label{Sec: Concl}

In this work, we propose the \rwff design framework to guide the design of CRDTs. \rwff leverages the remove-win strategy to resolve conflicting updates pertinent to remove operations, and provides generic design for a variety of data container types. Exemplar implementations over the Redis data type store show the effectiveness of \rwf. Performance measurements show the efficiency of CRDT implementations following \rwf.

In our future work, we will design more CRDTs using \rwf. We will also formally specify and verify the designs and implementations following \rwf. More comprehensive experimental evaluations under various workloads are also necessary.



\bibliographystyle{IEEEtran}
\bibliography{rwf-tr}

\newpage

\appendices

\section{\rwf-RPQ Design}

We design and implement a Replicated Priority Queue (RPQ), under the guidance of the Remove-Win Framework. The RPQ is a container of elements of the form $e = (id, priority)$. Each element is identified by its $id$, and without loss of generality, we assume that the priority value is an integer. The client can modify (the replica of) the RPQ by the following update operations:
\begin{itemize}
	\item \makebox[1.45cm]{$add(e,x)$\hfill}: enqueue element $e$ with initial priority $x$.
	\item \makebox[1.45cm]{$rmv(e)$\hfill}: remove the element $e$.
	\item \makebox[1.45cm]{$inc(e,\delta)$\hfill}: increase the priority of element $e$ by $\delta$ ($\delta$ may be negative).
\end{itemize}

\noindent Additionally, we assume that the RPQ supports the query operations below to better illustrate our RPQ design:
\begin{itemize}
	\item \makebox[1.7cm]{$empty()$\hfill}: returns $true$ if the RPQ is empty.
	\item \makebox[1.7cm]{$lookup(e)$\hfill}: returns $true$ if $e$ is in the RPQ.
	\item \makebox[1.7cm]{$get\_pri(e)$\hfill}: returns the priority value of $e$.
	\item \makebox[1.7cm]{$get\_max()$\hfill}: returns the $id$ and $priority$ of the element with the highest priority.
\end{itemize}

\noindent Following the \rwf-Skeleton, design of the RPQ is obtained by instantiating the \rwf-Skeleton and develop RPQ-specific stubs, as detailed below.

\subsection{RPQ Design}

Since conflicts concerning element existence is handled by the \rwf-Set, the user only needs to care about element values. The user needs to specify how priority values are initialized and updated by the RPQ APIs. More importantly, the user needs to develop conflict-resolving stubs and ``plug" them into the \rwf-Skeleton.

As for the $add$-$upd$ conflict, the priority value of an element $e$ is divided into two parts: the \textit{innate value} set by its initiating $add(e)$ operation, and the \textit{acquired value} updated by the following $inc(e,i)$ operations. In the RPQ design, the priority value exposed to the upper-layer application is the sum of innate and acquired values. The $add$ and $upd$ operations take effects on the innate and acquired values respectively and conflicts are prevented.

As for the $add$-$add$ conflict, the user needs to specify an total order among concurrent $add$ operations. This order decides the unique $add$ that finally ``wins", while other $add$s are overwritten. In our exemplar design, we can simply specify ``largest replica $id$ wins" (assuming that the $id$s of all replicas are totally ordered).

As for the $upd$-$upd$ conflict, there will be no this type of conflict in the priority queue case. It is because the add/subtraction of priority values (integers) naturally commute.



The detailed RPQ design is presented in Algorithm \ref{A:RWF-RPQ(p,q)} and Algorithm \ref{A:RWF-RPQ(u)}.

\begin{algorithm}
    \DontPrintSemicolon
    \caption{\rwf-RPQ (payloads and queries)}
    \label{A:RWF-RPQ(p,q)}
    \Payload $ E$: set of $(e, p_{ini})$ tuples, $T$: set of $(e, t)$ tuples, $V$: set of $(e,v_{inn}, v_{acq})$ tuples \; 
    \Initial $E = \emptyset, T = \emptyset, V = \emptyset$ \;
    \Query({$empty()$: boolean})
    {
        \Return $E \neq \emptyset$ \;
    }
    \Query({$lookup(e)$: boolean})
    {
        \Return $\exists p_{ini}:(e,p_{ini})\in E$ \;
    }
    \Query({$get\_pri(e)$: integer})
    {
        \Pre $lookup(e)$\;
        \Let $x,\delta: (e, x, \delta)\in V$\;
        \Return $x+\delta$\;
    }
    \Query({$get\_max()$: id, integer})
    {
        \Pre $\lnot{empty()}$\;
        \Let $e: lookup(e)\wedge \forall o: lookup(o) \wedge get\_pri(o)\leq get\_pri(e)$ \;
        \Return $e, get\_pri(e)$\;
    }
\end{algorithm}
\begin{algorithm}
	\DontPrintSemicolon
	\caption{\rwf-RPQ (updates)}
	\label{A:RWF-RPQ(u)}
    \Update({$ add(e,x) $})
    {
        \AtSource({$(e,x)$})
        {
            \Pre $\lnot{lookup(e)}$\;
            \Let $v^{rh} = t$ s.t. $(e,t) \in T$ \tcp*{$v^{rh} = \vec{0}$ if there is no $(e,t)$ in $T$.}
            \Let $p_{ini}$ be id of the initiator of this operation \;
        }
        \DownStream({$(e,x,p_{ini},v^{rh})$})
        {
            $rmv(e,v^{rh})$  \tcp*{Execute the \textbf{effect} part of $rmv(e)$ using $v^{rh}$.}
            \Let $pid : (e, pid )\in E$ \tcp*{$pid = -1$ if there is no $(e,pid)$ in $E$.} 
            \Let $t : (e, t )\in T$ \tcp*{$t = \vec{0}$ if there is no $(e,t)$ in $T$.}		
            \uIf(\tcp*[f]{Larger replica $id$ wins.}){$v^{rh} = t \ \wedge \ p_{ini}>pid$} 
            {
                $E:=E\setminus\{(e,pid)\}\cup \{(e,p_{ini})\}$\;
                \Let $x',\delta:(e,x',\delta)\in V$\tcp*{$x'=0$ and $\delta=0$ if there is no $(e,x',\delta)$ in $V$.}
                $V:=V\setminus\{(e,x',\delta)\}\cup \{(e,x,\delta)\}$\;				
            }
        }
    }
    \Update({$ inc(e,i) $}\tcp*[f]{$i\in \mathbb{Z}$, $i<0$ means `decrease'.})
    {
        \AtSource({$(e,i)$})
        {
            \Pre $lookup(e)$\;
            \Let $v^{rh} = t$ s.t. $(e,t) \in T$ \tcp*{$v^{rh} = \vec{0}$ if there is no $(e,t)$ in $T$.}
        }
        \DownStream({$(e,i,v^{rh} )$})
        {
            $rmv(e,v^{rh} )$  \tcp*{The same as the effect part of $add$.}
            \Let $t : (e, t )\in T$ \tcp*{$t = \vec{0}$ if there is no $(e,t)$ in $T$.} 		
            \uIf{$v^{rh} = t$}
            {
                \Let $x,\delta:(e,x,\delta)\in V$\tcp*{$x=0$ and $\delta=0$ if there is no $(e,x,\delta)$ in $V$.}
                $V:=V\setminus\{(e,x,\delta)\}\cup \{(e,x,\delta+i)\}$\;	
            }
        }
    }
    \Update({$rmv(e)$})
    {
        \AtSource({$(e)$})
        {
            \Pre $lookup(e)$\;
            \Let $v^{rh} = t$ s.t. $(e,t) \in T$ \tcp*{$v^{rh} = \vec{0}$ if there is no $(e,t)$ in $T$.}
            \Let $p_{ini}$ be id of the initiator of this operation \;
            $v^{rh}[p_{ini}] := v^{rh}[p_{ini}]+1$\;
        }
        \DownStream({$(e,v^{rh})$})
        {
            \Let $t:(e,t)\in T$\tcp*{$t = \vec{0}$ if there is no $(e,t)$ in $T$.}
            \uIf{$\exists k: t[k] < v^{rh}[k]$}
            {
                \Let $pid:(e,pid)\in E$\tcp*{$pid = -1$ if there is no $(e,pid)$ in $E$.}
                $E:=E\setminus\{(e,pid)\}$\;	
                \Let $x,\delta:(e,x,\delta)\in V$\tcp*{$x=0$ and $\delta=0$ if there is no $(e,x,\delta)$ in $V$.}
                $V:=V\setminus\{(e,x,\delta)\}$\;	
                \Let $t': \forall k:t'[k]:=\max(v^{rh}[k], t[k])$\;
                $T:=T\setminus\{(e,t)\}\cup\{(e,t')\}$\;
            }			
        }
    }
\end{algorithm}

\subsection{Illustrating Examples}

We use three examples to better illustrate the design of our RPQ. This first example mainly shows how the remove-win strategy works. The second example shows how the conflict resolution among non-remove operations within one phase works. The third example mainly shows that we don't need causal delivery for phases because we redo $rmv$ operations in non-remove operations using the rh-vec they carry.

\begin{figure}[ht]
    \centering
    \includegraphics[width=0.9\linewidth]{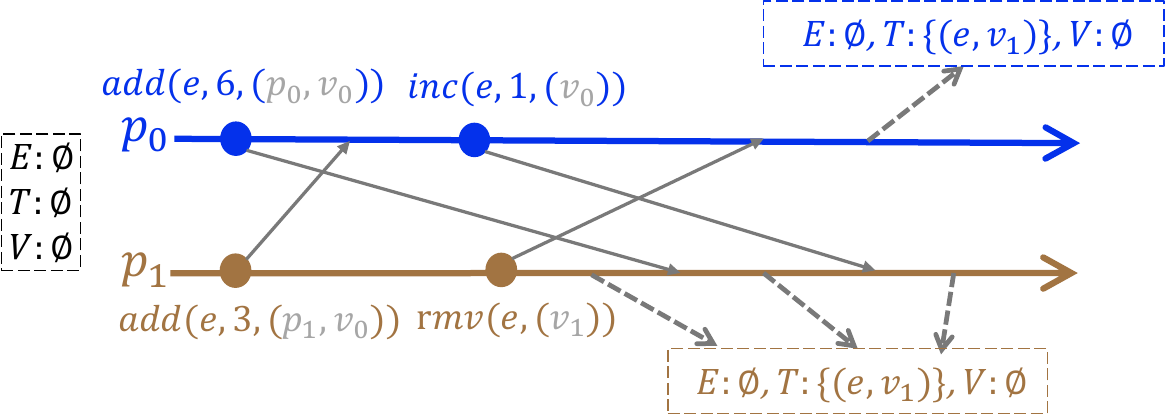}
    \caption{An example showing how an $rmv$ wins, where $v_0=[0,0]$, $v_1=[0,1]$.}
    \label{F:RW2}
\end{figure}

In the remove-win example in Figure \ref{F:RW2}, the $rmv$ operation initiated by $p_1$ is concurrent with the $add$ and $inc$ operations initiated by $p_0$. On $p_1$, after the $rmv$ operation is executed, the rh-vec of $e$ in $T$ is set to $v_1=[0,1]$, which is larger than the rh-vecs of $add$ and $inc$ on $p_0$. So when the remote events of $add$ and $inc$ arrives at $p_1$, they will be safely ignored, and the payload on $p_1$ remains unchanged whether $add$ and $inc$ arrive or not. When the remote event of $rmv$ from $p_1$ is received by $p_0$, $p_0$ will remove the element $e$ from $E$, since the $rmv$ carries the larger rh-vec $v_1$.

In the example of conflict resolution among non-remove operations in Figure \ref{F:RW1}, the payloads of $p_0$ and $p_1$ are initially empty. First, we have $p_0$ and $p_1$ add the element $e$ concurrently, with the same rh-vec $v_0 = [0, 0]$. This indicates that they belong to the same phase and need conflict resolution. Here we adopt the strategy that ``larger replica id wins". Thus the $add$ of $p_1$ wins. We find that the tuple in $E$ on $p_0$ remains $(e,p_0)$ until it finally receives the $add$ operation from $p_1$ and the tuple in $E$ is changed to $(e, p_1)$. Then we have $p_0$ and $p_1$ increase $e$ with the rh-vec $v_0$, and the increased values merged without conflict into the acquired value of $e$. Finally $p_0$ and $p_1$ converge to the same state.

\begin{figure}[ht]
    \centering
    \includegraphics[width=0.9\linewidth]{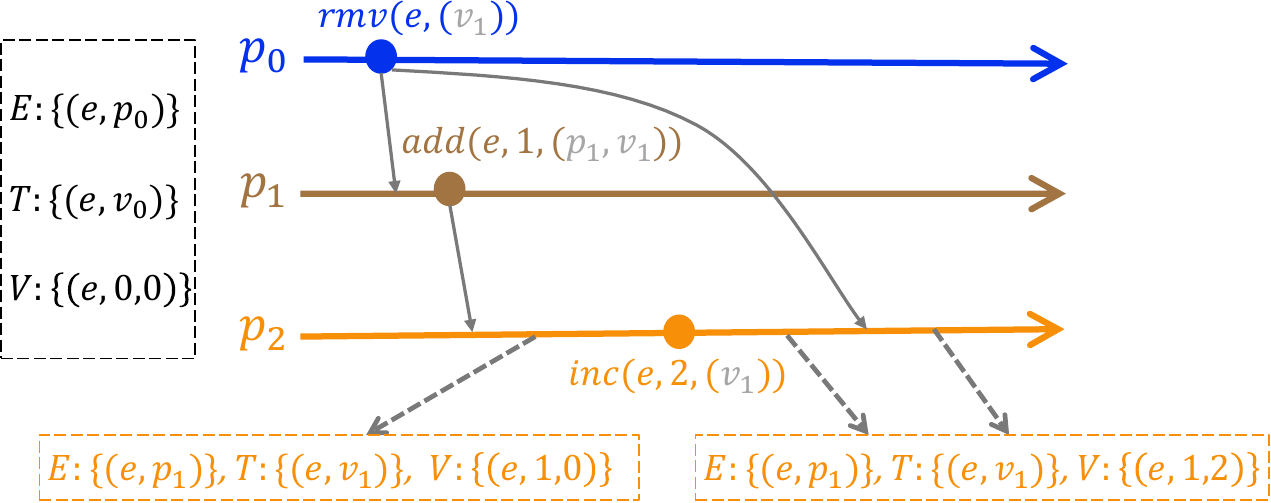}
    \caption{No need for causal delivery for phase, where $v_0=[0,0,0]$, $v_1=[1,0,0]$.}
    \label{F:RW3}
\end{figure}

In the example in Figure \ref{F:RW3}, we show the reason why we don't need causal delivery. The $rmv$ initiated by $p_0$ is visible to the $inc$ initiated by $p_2$, not directly but via the $add$ operation initiated by $p_1$. The rh-vec is initially $v_0 = [0,0,0]$. The $rmv$ on $p_0$ updates the rh-vec to $v_1 = [1,0,0]$. Then $v_1$ is transmitted to from $p_0$ to $p_1$ and from $p_1$ to $p_2$, and the missing $rmv$ operation is redone at $p_2$, updating the rh-vec of $p_2$ to $v_1$. Thus when the $rmv$ operations arrives late at $p_2$ (bringing with it the rh-vec $v_1$), it will be safely ignored since $p_2$ has already obtained the rh-vec $v_1$ before. Without the redo of the $rmv$ triggered by $add$ that update the rh-vec on $p_2$, the $rmv$ from $p_0$ will arrive at $p_2$ late and falsely removes element $e$. Causal message delivery is necessary to ensure that on $p_2$, $rmv$ is delivered before $add$.

\begin{figure}[ht]
    \centering
    \includegraphics[width=0.9\linewidth]{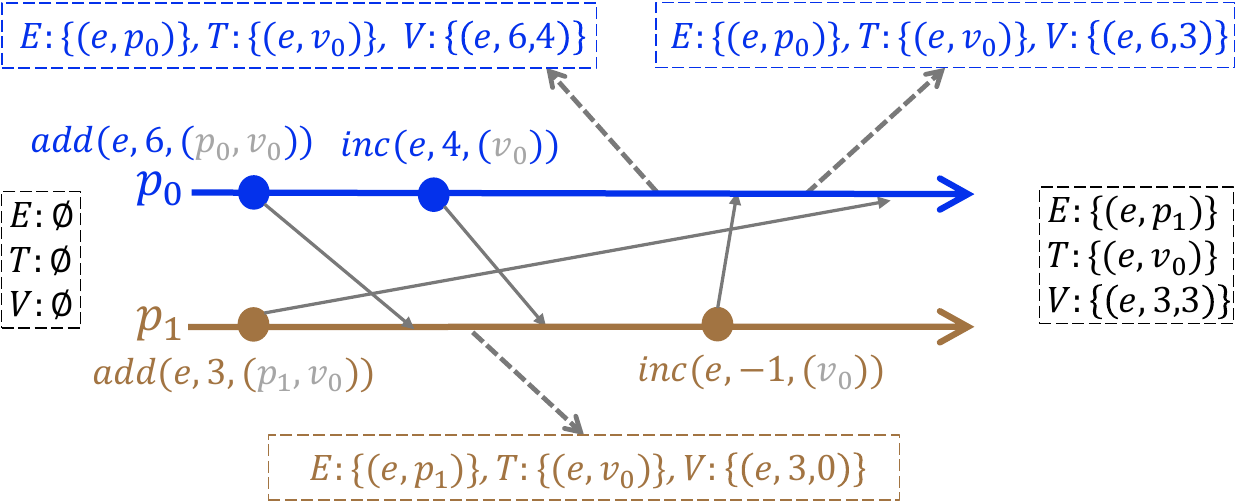}
    \caption{Conflict resolution among non-remove operations, where $v_0=[0,0]$.}
    \label{F:RW1}
\end{figure}

\section{Remove-Win RPQ Design}

Here we try to design a Remove-Win RPQ without our \rwf-Skeleton. 

Note that the classic remove-win doesn't mean that the remove operation simply kills all other concurrent non-remove operations. Sometimes these concurrent non-remove operations will still take effect. See the example in figure \ref{F:RW1}. There are no communication between two processes. Therefor $r_1$ and $a_1$ are concurrent with $r_2$ and $a_2$. Although both $a_1$ and $a_2$ has a concurrent remove operation ($r_2$ and $r_1$) that may kill them due to the remove-win semantics, combined they win over the remove operations. Then the element is in the RPQ rather than removed. This is reasonable, because if you linearize the causal order of these four operations, the last operation will always be an add.





\begin{figure}[ht]
	\centering
	\includegraphics[width=0.3\linewidth]{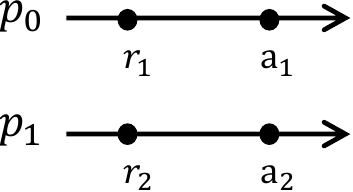}
	\caption{The case of remove-win.}
	\label{F:RW1}
\end{figure}

The detailed design is shown in Algorithm \ref{A:Remove-Win-RPQ}. Here we assume that causal delivery is provided by the underlying network. We can use $now()$ function to get the vector clock of the current operation. We denote $v_1 \parallel v_2$ as two vector clocks $v_1$ and $v_2$ are parallel, and $v_1 < v_2$ means $v_1$ is less than $v_2$. Note that this vector clock indicates the visible relation between operations: $op_1\vis op_2 \iff op_1.vec < op_2.vec$.

We first discuss the existence of elements. Firstly the Remove-Win specification: $e\in RPQ \iff \exists add(e) \wedge \forall rmv(e). \exists add(e). rmv(e) \vis add(e)$. We notice that to decide if an element $e$ is in the RPQ, we only need to store all the $add(e)$ and $rmv(e)$ operations that may be effective, which means there is no $add(e)$ or $rmv(e)$ operation that happen after them. Then we decide if the element $e$ is in the RPQ strictly by the Remove-Win specification.

Then the value of elements, we resolve the conflicts of initial value brought by concurrent $add$ operations with the process id. Here we let the $add$ with larger process id win, and yet we store all the value records of these $add$ operations. As for the $inc$ operations, we let it only increase the value records brought by the $add$ operations that are visible to it. We use the vector clock to identify this. Because of causal delivery, the $inc$ operation will be correctly applied at all replicas.

\begin{algorithm}[htbp]
	\DontPrintSemicolon
	\caption{Remove-Win RPQ}
	\label{A:Remove-Win-RPQ}
	\Payload $A$: set of $(e,t,id,x,\delta)$ tuples, $R$: set of $(e,t)$ tuples \;
	\Initial $A = \emptyset, R = \emptyset$ \; 
	\Query({$empty()$: boolean})
	{
		\Return $\forall e: (e,t,id,x,\delta)\in A \rightarrow \lnot lookup(e)$ \;
	}
	\Query({$lookup(e)$: boolean})
	{
		\Return $\exists t: (e,t,id,x,\delta)\in A \wedge \nexists t': (e,t')\in R$ \;
	}
	\Query({$get\_pri(e)$: integer})
	{
		\Pre $lookup(e)$\;
		\Let $id,x,\delta: \forall (e,t,id',x',\delta')\in A: id'<id$\;
		\Return $x+\delta$\;
	}
	\Query({$get\_max()$: id, integer})
	{
		\Pre $\lnot{empty()}$\;
		\Let $e: lookup(e)\wedge \forall o: lookup(o) \wedge get\_pri(o)\leq get\_pri(e)$ \;
		\Return $e, get\_pri(e)$\;
	}
	\Update({$ add(e,x) $})
	{
		\AtSource({$(e,x)$})
		{
			\Pre $\lnot lookup(e)$\;
			\Let $t=now()$\;
			\Let $id$: the id of the process \;
		}
		\DownStream({$(e, x, t, id)$})
		{
			$A:=A\cup\{(e,x,t,id,0)\}$\;
			\ForEach{$(e,x,t,id,\delta)\in A$}
			{
				\lIf{$t'<t$}{$A:=A\setminus\{(e,x,t,id,\delta)\}$}
			}
			\ForEach{$(e,t')\in R$}
			{
				\lIf{$t'<t$}{$R:=R\setminus\{(e,t')\}$}
			}	
		}
	}
	\Update({$rmv(e)$})
	{
		\AtSource({$(e)$})
		{
			\Pre $lookup(e)$\;
			\Let $t=now()$\;
		}
		\DownStream({$(e, t)$})
		{
			$R:=R\cup\{(e,t)\}$\;
			\ForEach{$(e,x,t,id,\delta)\in A$}
			{
				\lIf{$t'<t$}{$A:=A\setminus\{(e,x,t,id,\delta)\}$}
			}
		}
	}
	\Update({$ inc(e,i) $})
	{
		\AtSource({$(e,i)$})
		{
			\Pre $lookup(e)$\;
			\Let $t=now()$\;
		}
		\DownStream({$(e, i, t)$})
		{
			\ForEach{$(e,x,t',id,\delta)\in A$}
			{
				\lIf{$t'<t$}{$A:=A\setminus\{(e,x,t',id,\delta)\}\cup \{(e,x,t',id,\delta+i)\}$}
			}	
		}
	}

\end{algorithm}

\section{\rwf-List Design}

We design and implement a Replicated List under the guidance of the Remove-Win Framework. The List is a container of elements of the form $e = (id, content, properties)$. Elements are totally ordered.
An element has its unique ID, the content (letter, word, or paragraph...), and properties (font, size, color, shape...). The content of one element will not be changed in co-editing scenario. Clients can modify the list by the following update operations:
\begin{itemize}
	\item \makebox[1.95cm]{$add(e,e_{p},P)$\hfill}: add the element $e$ after $e_p$ with initial properties $P$.
	\item \makebox[1.95cm]{$upd(e,p)$\hfill}: update the element $e$ with some new property $p$.
	\item \makebox[1.95cm]{$rmv(e)$\hfill}: remove the element $e$.
\end{itemize}

\noindent Additionally, we assume that the List supports the query operations below:
\begin{itemize}
	\item \makebox[2.1cm]{$empty()$\hfill}: returns $true$ if the List is empty.
	\item \makebox[2.1cm]{$lookup(e)$\hfill}: returns $true$ if $e$ is in the List.
	\item \makebox[2.1cm]{$properties(e)$\hfill}: returns the properties of $e$.
	\item \makebox[2.1cm]{$read\_list()$\hfill}: returns the list of elements with its content and properties, totally ordered.
\end{itemize}

\noindent Following the \rwf-Skeleton, design of the \rwf-List is obtained by instantiating the \rwf-Skeleton and develop List-specific stubs.

The detailed \rwf-List design is presented in Algorithm \ref{A:RWF-List(p,q)} and Algorithm \ref{A:RWF-List(u)}.

Here we use the Logoot ID to identify the position of the element. The Logoot ID is unique, totally ordered and dense. Hence the list is transformed into the ordered set whose elements are ordered by the Logoot ID. By using the \rwf-Skeleton, the existence of elements is properly handled. The order of elements is identified by Logoot IDs. Now we only need to care about the consistence of values of elements.

Moreover, the innate value of elements brought by $add$ operations are handled by the \rwf-Skeleton. The $add$-$add$ conflict resolution is done by using the $pid$ of the initiating process. As for the $add$-$upd$ conflict, we let the $update$ operations win over $add$ operations if they are in the same phase. As for the $upd$-$upd$ conflict, we attach a totally-ordered lamport-clock generated by $now()$ function to each $update$ operation. Then we adopt the last-write-win policy for conflicting $update$ operations in the same phase.

\begin{algorithm}
	\DontPrintSemicolon
	\caption{\rwf-List (payloads and queries)}
	\label{A:RWF-List(p,q)}
	\Payload $E$: set of $(e, p_{ini}, pos)$ tuples, $T$: set of $(e, t)$ tuples, $V$: set of $(e,I, A)$ tuples  \tcp*{$pos$: Logoot ID, $I$: set of $property_{inn}$, $A$: set of $(property_{acc}, t)$ tuples }
	\Initial $E = \emptyset, T = \emptyset, V = \emptyset$ \;
	\Query({$empty()$: boolean})
	{
		\Return $\nexists e: lookup(e)$ \;
	}
	\Query({$lookup(e)$: boolean})
	{
		\Return $\exists p_{ini}:(e, p_{ini}, pos)\in E \wedge p_{ini}\neq -1 \wedge \nexists A:(e,\emptyset,A)\in V$ \;
	}
	\Query({$properties(e)$: properties})
	{
		\Pre $lookup(e)$\;
		\Let $I,A: (e, I,A)\in V$\;
		\Return for each kind of property, the value in $A$ with max $t$, or the value in $I$ if no such property in $A$\;
	}
	\Query({$read\_list()$: list})
	{
		\Pre $\lnot{empty()}$\;
		\Let $R={(e,pos)|(e, p_{ini}, pos)\in E \wedge lookup(e)}$\;
		\Return the list of $e$ in $R$, sorted by $pos$\;
	}
\end{algorithm}
\begin{algorithm}
	\DontPrintSemicolon
	\caption{\rwf-List (updates)}
	\label{A:RWF-List(u)}
	\Update({$ add(e,e_{p},P) $}\tcp*[f]{add $e$ after $e_{p}$, or at the beginning if $e_{p}=null$, $P$: initial properties})
	{
		\AtSource({$(e,e_{p},P)$})
		{
			\Pre $\lnot{lookup(e)} \wedge( lookup(e_{p}) \vee e_{p}\ is\ null)$\;
			\Let $v^{rh} = t$ s.t. $(e,t) \in T$ \tcp*{$v^{rh} = \vec{0}$ if there is no $(e,t)$ in $T$.}
			\Let $p_{ini}$ be id of the initiator of this operation \;
			\Let $pos: (e, p_{ini}, pos)\in E$ if there is such tuple, or otherwise the proper Logoot ID after $e_{p}$ and before the next element of $e_{p}$\;
		}
		\DownStream({$(e,pos,P ,p_{ini},v^{rh})$})
		{
			$rmv(e,v^{rh})$  \tcp*{Execute the \textbf{effect} part of $rmv(e)$ using $v^{rh}$.}
			\Let $pid : (e, pid,pos )\in E$ \tcp*{$pid = -1$ if there is no $(e,pid,pos)$ in $E$.} 
			\Let $t : (e, t )\in T$ \tcp*{$t = \vec{0}$ if there is no $(e,t)$ in $T$.}		
			\uIf(\tcp*[f]{Larger replica $id$ wins.}){$v^{rh} = t \ \wedge \ p_{ini}>pid$} 
			{
				$E:=E\setminus\{(e,pid,pos)\}\cup \{(e,p_{ini},pos)\}$\;
				\Let $(e,I, A)\in V$\tcp*{$I=\emptyset$ and $A=\emptyset$ if there is no $(e,I, A)$ in $V$.}
				$V:=V\setminus\{(e,I,A)\}\cup \{(e,P,A)\}$\;				
			}
		}
	}
	\Update({$ upd(e,p) $}\tcp*[f]{$p$ is some property})
	{
		\AtSource({$(e,p)$})
		{
			\Pre $lookup(e)$\;
			\Let $v^{rh} = t$ s.t. $(e,t) \in T$ \tcp*{$v^{rh} = \vec{0}$ if there is no $(e,t)$ in $T$.}
			\Let $t_u=now()$\tcp*{lamport clock}
		}
		\DownStream({$(e,p,t_u,v^{rh} )$})
		{
			$rmv(e,v^{rh} )$  \tcp*{The same as the effect part of $add$.}
			\Let $t : (e, t )\in T$ \tcp*{$t = \vec{0}$ if there is no $(e,t)$ in $T$.} 		
			\uIf{$v^{rh} = t$}
			{
				\Let $(e,I, A)\in V$\tcp*{$I=\emptyset$ and $A=\emptyset$ if there is no $(e,I, A)$ in $V$.}
				$V:=V\setminus\{(e,I, A)\}\cup \{(e,I, A\cup{(p,t_u)})\}$\;	
			}
		}
	}
	\Update({$rmv(e)$})
	{
		\AtSource({$(e)$})
		{
			\Pre $lookup(e)$\;
			\Let $v^{rh} = t$ s.t. $(e,t) \in T$ \tcp*{$v^{rh} = \vec{0}$ if there is no $(e,t)$ in $T$.}
			\Let $p_{ini}$ be id of the initiator of this operation \;
			$v^{rh}[p_{ini}] := v^{rh}[p_{ini}]+1$\;
		}
		\DownStream({$(e,v^{rh})$})
		{
			\Let $t:(e,t)\in T$\tcp*{$t = \vec{0}$ if there is no $(e,t)$ in $T$.}
			\uIf{$\exists k: t[k] < v^{rh}[k]$}
			{
				\Let $pid:(e,pid,pos)\in E$\tcp*{$pid = -1$ if there is no $(e,pid,pos)$ in $E$.}
				$E:=E\setminus\{(e,pid,pos)\}\cup \{(e,-1,pos)\}$\;	
				\Let $(e,I, A)\in V$\tcp*{$I=\emptyset$ and $A=\emptyset$ if there is no $(e,I, A)$ in $V$.}
				$V:=V\setminus\{(e,I,A)\}$\;	
				\Let $t': \forall k:t'[k]:=\max(v^{rh}[k], t[k])$\;
				$T:=T\setminus\{(e,t)\}\cup\{(e,t')\}$\;
			}			
		}
	}
\end{algorithm}
\section{Remove-Win List Design}

Here we try to design a Remove-Win List without our \rwf-Skeleton. The detailed design is shown in Algorithm \ref{A:Remove-Win-List}. The same as Remove-Win RPQ, here we assume that causal delivery is provided by the underlying network. And we can use $now()$ function to get the vector clock of the current operation. Like \rwf-List, we use Logoot ID to identify the position of an element in the list. Then the consistency of element order is guaranteed.

We use the same technique of the Remove-Win RPQ to ensure the consistency of the existence of elements and the remove-win semantics, which is to store the effective $add$ and $rmv$ operations, and then decide if the element $e$ is in the list by the Remove-Win specification.

Then the consistency of the element value. We store all the initial value brought by $add$ operations that are still effective, together with the process id of the replica that generated the $add$, as value records. The $update$ operations, like it is in Remove-Win RPQ, will update all the value records of $add$ operations that is visible to it. The $update$ operations adopt a last-write-win strategy if two $update$ want to update the same value record simultaneously. Finally, the value record that is read by clients is that with the highest process id.

\begin{algorithm}[htbp]
	\DontPrintSemicolon
	\caption{Remove-Win List}
	\label{A:Remove-Win-List}
	\Payload $L$: set of $(e,pos)$ tuples, $A$: set of $(e,t,id,P)$ tuples, $R$: set of $(e,t)$ tuples \tcp*{$P$: set of (property, t, id) tuples }
	\Initial $L = \emptyset, A = \emptyset, R = \emptyset$ \; 
	\Query({$empty()$: boolean})
	{
		\Return $\nexists e: lookup(e)$ \;
	}
	\Query({$lookup(e)$: boolean})
	{
		\Return $\exists t: (e,t,id,P)\in A \wedge \nexists t': (e,t')\in R$ \;
	}
	\Query({$properties(e)$: properties})
	{
		\Pre $lookup(e)$\;
		\Let $id,P: \forall (e,t,id',P')\in A: id'<id$\;
		\Return properties in $P$\;
	}
	\Query({$read\_list()$: list})
	{
		\Pre $\lnot{empty()}$\;
		\Let $R={(e,pos)|(e, pos)\in L \wedge lookup(e)}$\;
		\Return the list of $e$ in $R$, sorted by $pos$\;
	}
	\Update({$ add(e,e_{p},P) $}\tcp*[f]{add $e$ after $e_{p}$, or at the beginning if $e_{p}=null$, $P$: initial properties})
	{
		\AtSource({$(e,e_{p},P)$})
		{
			\Pre $\lnot{lookup(e)} \wedge( lookup(e_{p}) \vee e_{p}\ is\ null)$\;
			\Let $t=now()$\;
			\Let $id$: the id of the process \;
			\Let $pos: (e,pos)\in L$ if there is such tuple in $L$, or otherwise the proper logoot ID after $e_{p}$ and before the next element of $e_{p}$\;
		}
		\DownStream({$(e, P, t, id, pos)$})
		{
			$L:=L\cup\{(e,pos)\}$\;
			$A:=A\cup\{(e,t,id,P\times\{(t,id)\})\}$\;
			\ForEach{$(e,t,id,P)\in A$}
			{
				\lIf{$t'<t$}{$A:=A\setminus\{(e,t,id,P)\}$}
			}
			\ForEach{$(e,t')\in R$}
			{
				\lIf{$t'<t$}{$R:=R\setminus\{(e,t')\}$}
			}	
		}
	}
	\Update({$rmv(e)$})
	{
		\AtSource({$(e)$})
		{
			\Pre $lookup(e)$\;
			\Let $t=now()$\;
		}
		\DownStream({$(e, t)$})
		{
			$R:=R\cup\{(e,t)\}$\;
			\ForEach{$(e,t,id,P)\in A$}
			{
				\lIf{$t'<t$}{$A:=A\setminus\{(e,t,id,P)\}$}
			}
		}
	}
	\Update({$ upd(e,p) $}\tcp*[f]{$p$ is some property})
	{
		\AtSource({$(e,p)$})
		{
			\Pre $lookup(e)$\;
			\Let $t=now()$\;
			\Let $id$: the id of the process \;
		}
		\DownStream({$(e, p, t, id)$})
		{
			\ForEach{$(e,t',id',P)\in A$}
			{
				\If{$t'<t$}
				{
					\Let $p',t_u,id_u: (p',t_u,id_u)\in P$ and $p'$ is the same type of $p$\;
					\lIf{$t_u<t \vee (t_u \parallel t \wedge id_u<id)$}
					{$P:=P\setminus\{(p',t_u,id_u)\}\cup\{(p,t,id)\}$}
				}
			}	
		}
	}
	
\end{algorithm}

\section{Experiment Result}

In this section, we provide more evaluation results and discussions.

\subsection{RPQ max difference between replicas}

\begin{figure}[h]
    \centering
    \includegraphics[width=\linewidth]{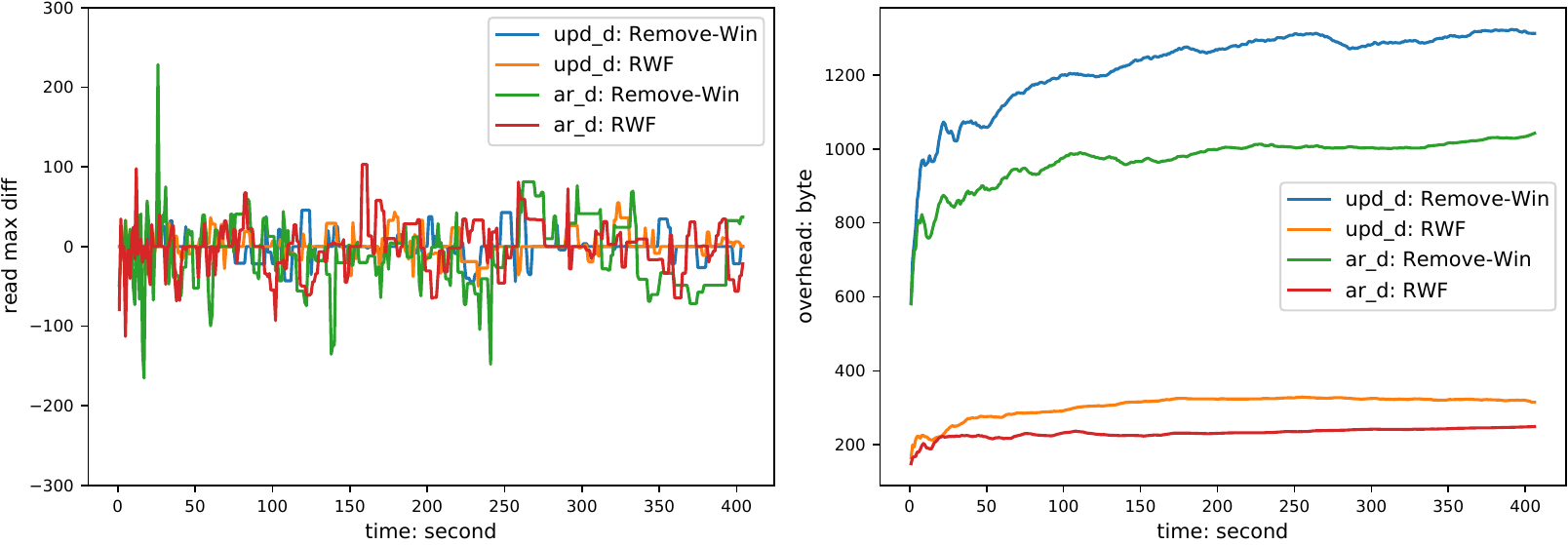}
    \caption{The performance of RPQs. Compare max read form different servers at the same time.}
    \label{F:exp-rpq-cmp}
\end{figure}

Here we compare the max read from two different replicas at the same time. The experiment settings are the same with the previous RPQ experiment. The statistics are shown in Table \ref{T: Consis-app}, and the result is shown in Fig. \ref{F:exp-rpq-cmp}.

The results are principally the same with those by comparing the replicated queue and the linearized queue. The difference vibrates mostly between -100 and 100. And the $add$/$rmv$-dominant workload pattern causes more differences. As for metadata overhead, it slowly increases, the $add$/$rmv$-dominant pattern causes higher overhead, and the \rwf-RPQ has less metadata overhead than the Remove-Win RPQ. The reasons are discussed in the previous sections. 

\begin{table}[tbp]
    \centering
    \caption{Data inconsistency on average.} 
    \label{T: Consis-app}
    \begin{tabu} to .6\linewidth {X[1.2,m,l]*2{X[1,m,l]}}
        \toprule      
        & \multicolumn{2}{l}{RPQ-replica {\scriptsize (Fig.\ref{F:exp-rpq-cmp})}} \\
        \cmidrule{2-3}
        & r & rwf \\
        \midrule
        {\small $upd$-dom} & 8.7 & 9.2    \\
        {\small $a/r$-dom} & 33.2 & 21.4  \\
        \bottomrule    
    \end{tabu}
\end{table}

\subsection{Impact of Concurrency among Operations}

There are three environment factors we can tune to control the impact of concurrency among operations. Thus, we conduct three experiments accordingly, tuning one factor in each experiment. Specifically, to control the concurrency among operations in the time dimension, we tune the speed at which operations are issued from clients to the servers. We increase the operation speed from 500 to 10,000 ops/s for RPQ, and from 50 to 1,000 ops/s for list. To control the concurrency in the space dimension, we change the network delay and the number of replicas. We tune the inter-data center delay from \nordis{20}{4}ms to \nordis{380}{76}ms, and tune the intra-data center delay from \nordis{4}{0.8}ms to \nordis{76}{15.2}ms. As for the number of replicas, we increase the number of Redis instances from 1 to 5 in every data center, and fix the operation generation speed for each Redis instance.

After the discussion of the former experiment, we here focus on comparing the max value between server and local record for RPQ, and comparing the list edit distance between lists read from two replicas.

\begin{figure}[h]
    \centering
    \includegraphics[width=.9\linewidth]{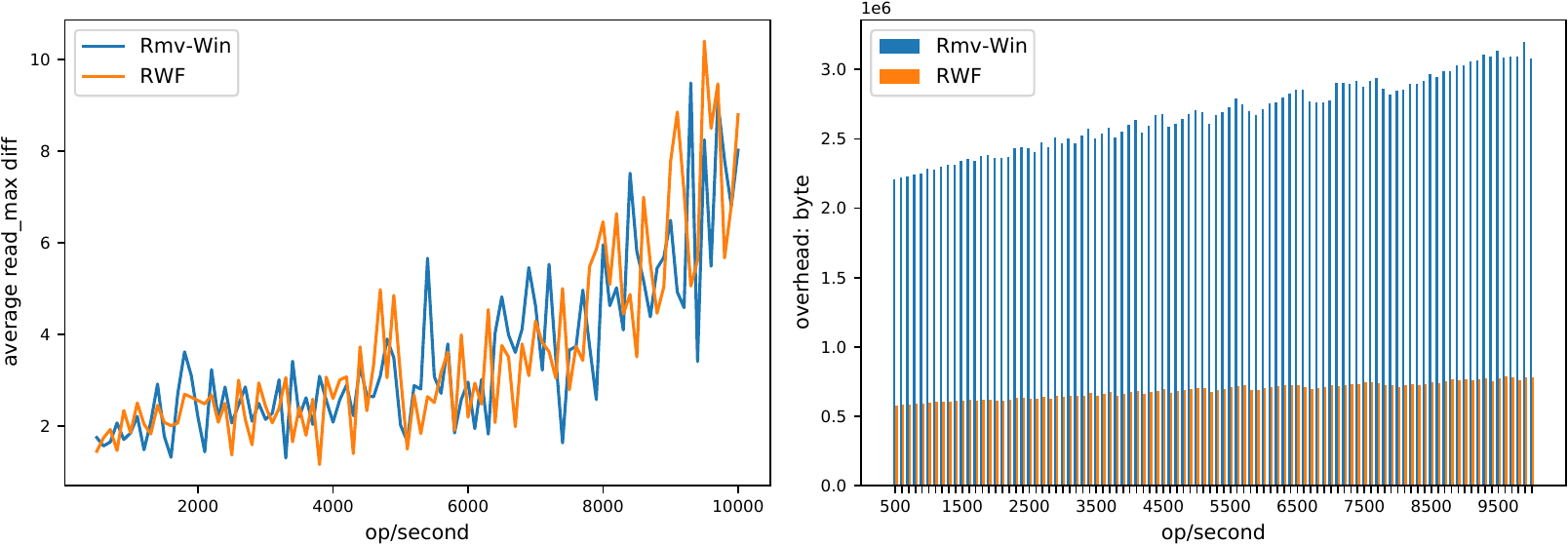}
    \caption{The performance of RPQs over different operation speed.}
    \label{F:exp-rpq-speed}
\end{figure}

\begin{figure}[h]
    \centering
    \includegraphics[width=.9\linewidth]{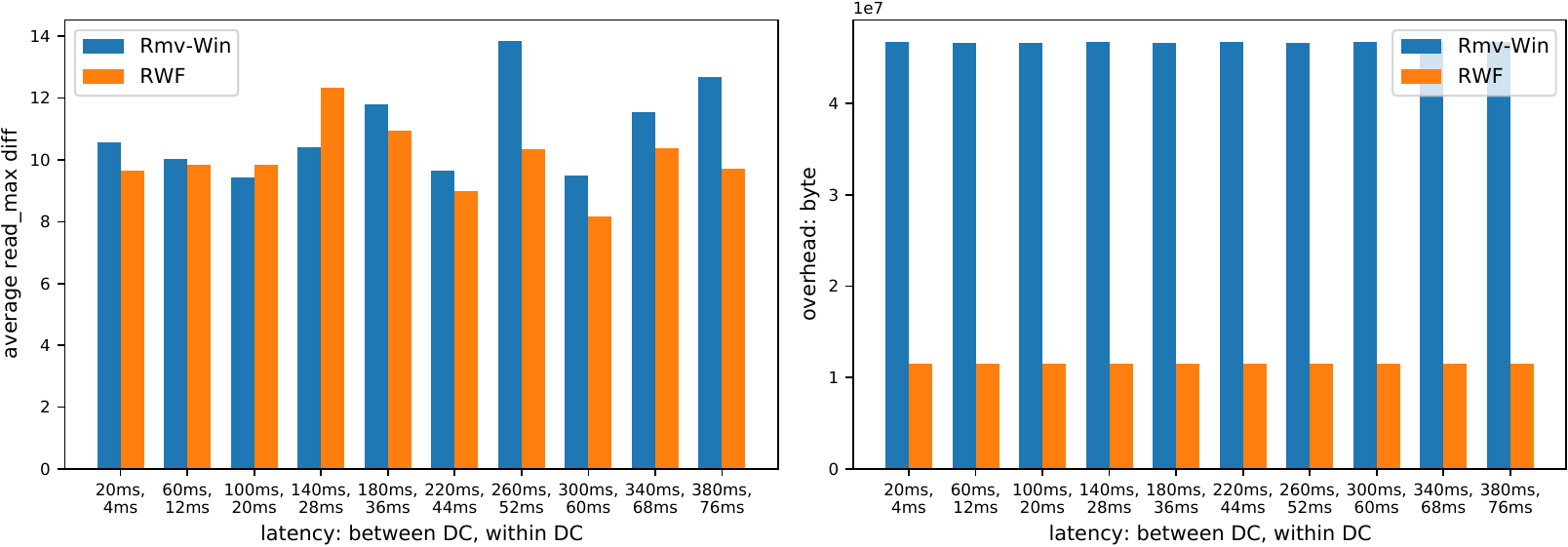}
    \caption{The performance of RPQs over different network delay.}
    \label{F:exp-rpq-delay}
\end{figure}

\begin{figure}[h]
    \centering
    \includegraphics[width=.9\linewidth]{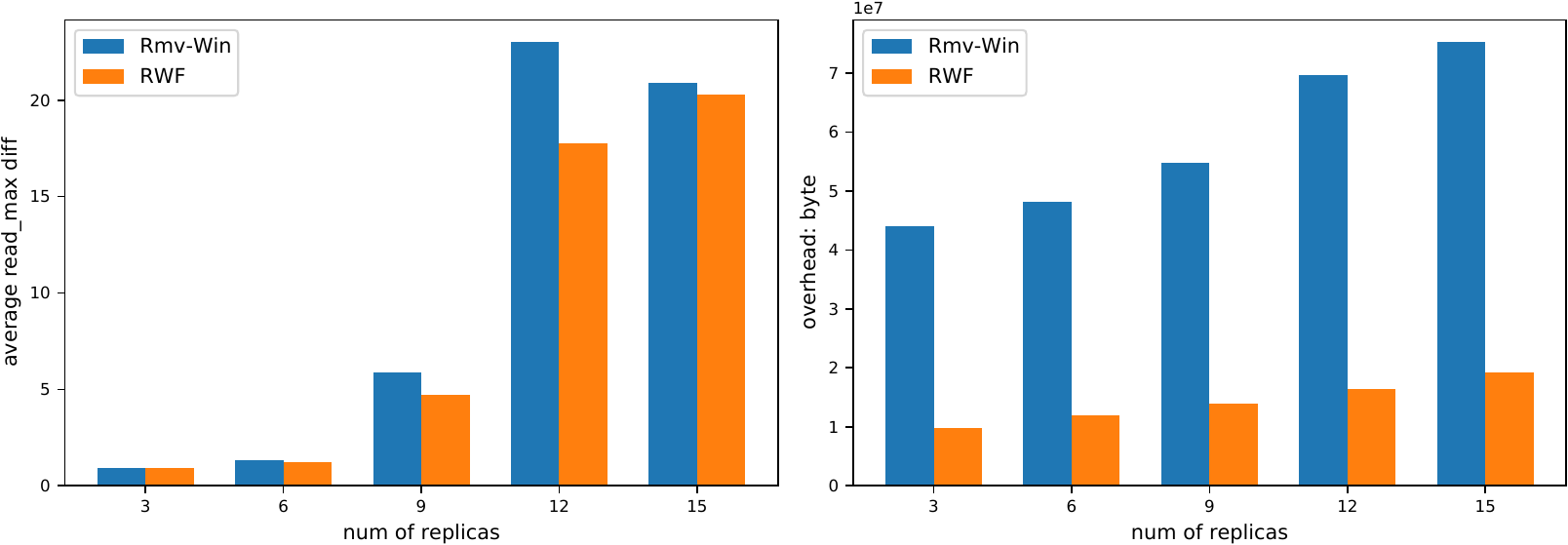}
    \caption{The performance of RPQs over different number of replicas.}
    \label{F:exp-rpq-replica}
\end{figure}

\begin{figure}[h]
    \centering
    \includegraphics[width=.9\linewidth]{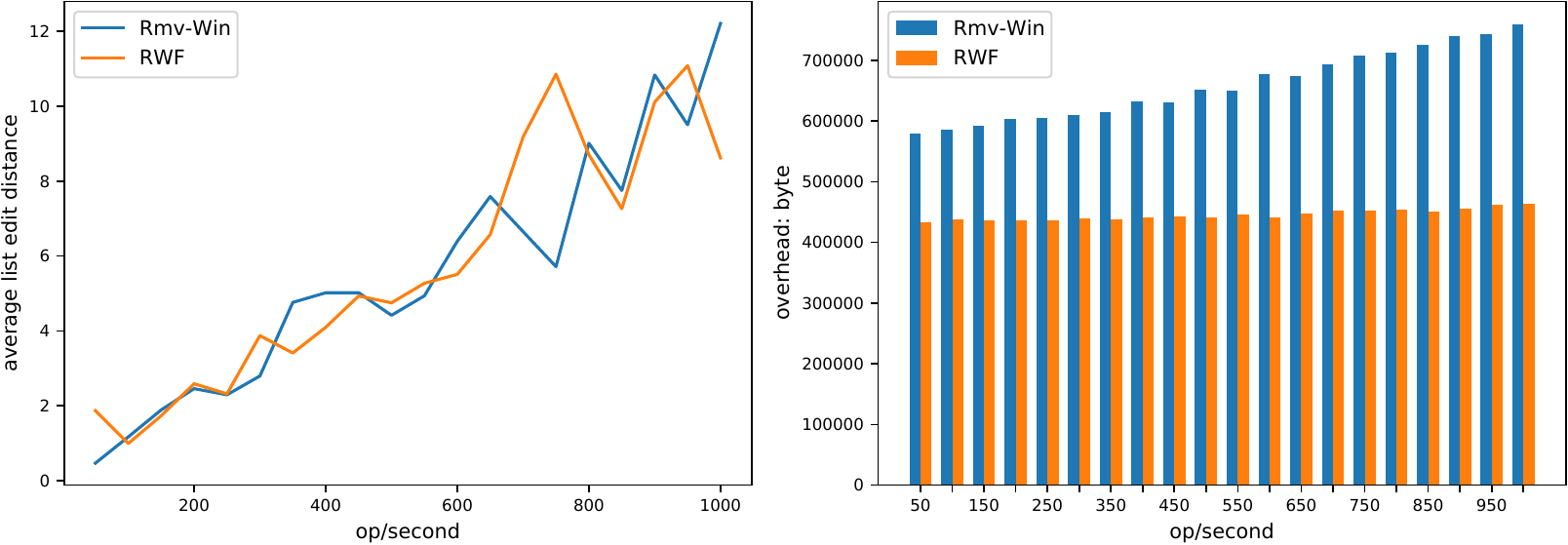}
    \caption{The performance of Lists over different operation speed.}
    \label{F:exp-list-speed}
\end{figure}

\begin{figure}[h]
    \centering
    \includegraphics[width=.9\linewidth]{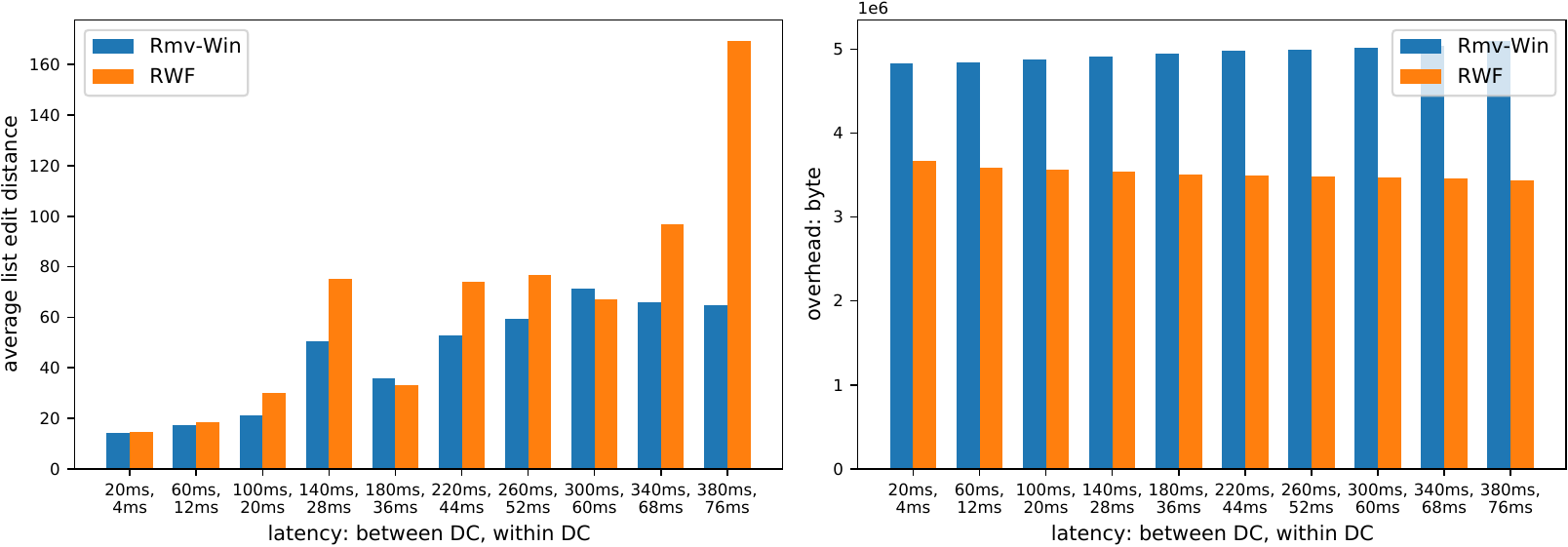}
    \caption{The performance of Lists over different network delay.}
    \label{F:exp-list-delay}
\end{figure}

\begin{figure}[h]
    \centering
    \includegraphics[width=.9\linewidth]{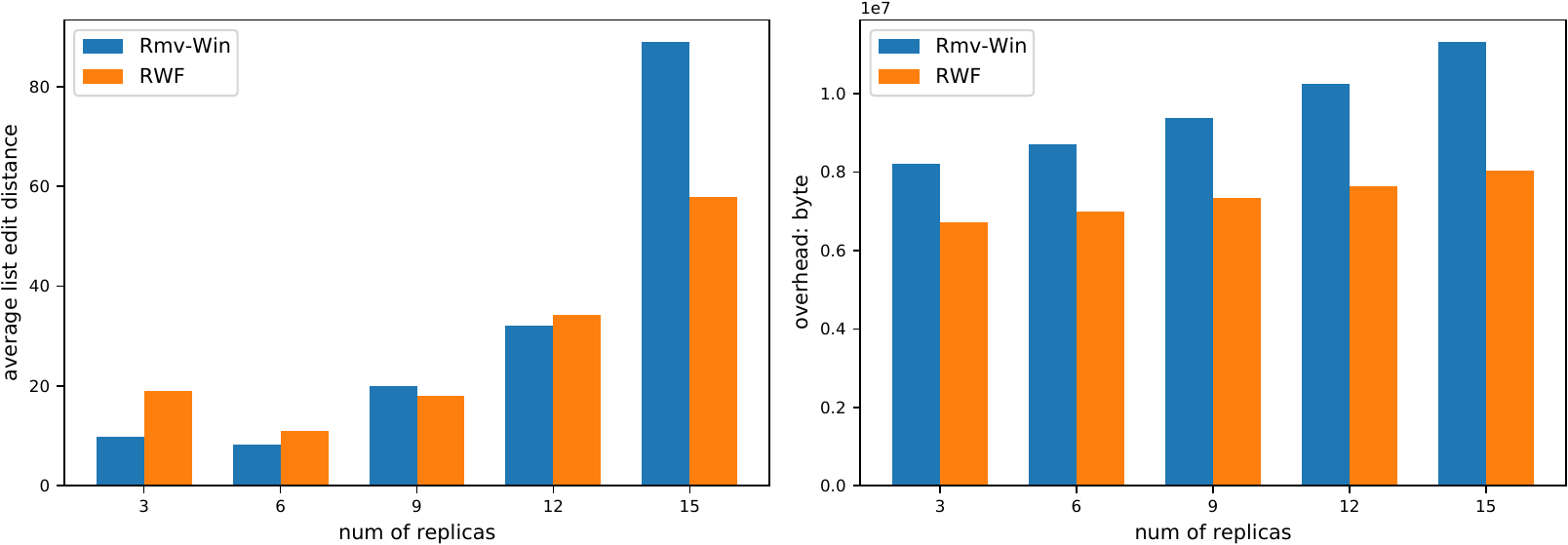}
    \caption{The performance of Lists over different number of replicas.}
    \label{F:exp-list-replica}
\end{figure}

As for the data consistency, we find that the average error $\bar x$ of $read\_max$ for both RPQs, and the list edit distance for both lists increases linearly with the concurrency among operations, as shown in Fig. \ref{F:exp-rpq-speed}, \ref{F:exp-rpq-delay} and \ref{F:exp-rpq-replica} for RPQ, and Fig. \ref{F:exp-list-speed}, \ref{F:exp-list-delay} and \ref{F:exp-list-replica} for List. This is mainly because the CRDT guarantees strong eventual consistency, and the inconsistency is mainly determined by the number of operations that are yet to be synchronized. As the concurrency among operations increases, the number of operations to be synchronized increases linearly. Thus we have the read differences increase linearly.

As for the metadata overhead, at the end of each run of the experiment, we measure the average total metadata overhead during this run. We find that the Remove-Win CRDTs have more metadata overhead as the operation speed increases, as shown in Fig. \ref{F:exp-rpq-speed} and \ref{F:exp-list-speed}. This is because the Remove-Win CRDTs require causal delivery. And as the operation speed increases, there are more causally unready operations that need more memory to deal with. And our RWF CRDTs do not need to deal with causally unready operations. They do not require causal delivery. As long as the number of operations conducted on the queue is statistically similar, the metadata overhead is also similar.

The message delay has less impact on the data consistency and the metadata overhead, as shown in Fig. \ref{F:exp-rpq-delay} and \ref{F:exp-list-delay}. We think this is because the message delay has less influence on the concurrency among operations than the other two factors in our experiment setups.

The metadata overhead of our CRDTs increases as there are more replicas in Fig. \ref{F:exp-rpq-replica} and \ref{F:exp-list-replica}. Not only because the concurrency among operations increases as the number of replica increases, since we fix the operation generation speed for each replica, but also the dimension of both vector clock and rh-vec get increased, as they are equal to the number of replicas on the server side. Thus the metadata overhead (for recording the vector) increases linearly as the number of replicas increases.

\end{document}